\documentclass[prb,twocolumn,aps,superscriptaddress,nofootinbib]{revtex4-2}

\usepackage{graphicx}
\usepackage{dcolumn}
\usepackage{bm}
\usepackage{amssymb}
\usepackage{amsmath}
\usepackage{physics}
\usepackage{sublabel}
\usepackage[utf8]{inputenc}
\usepackage{hyperref}
\usepackage[usenames, dvipsnames]{color}
\usepackage[english]{babel}
\usepackage[T1]{fontenc}
\usepackage{bbm}

\usepackage{float}
\usepackage{verbatim}
\usepackage[caption=false]{subfig}
\usepackage{xfrac}
\usepackage{upgreek}
\hypersetup{colorlinks = true, linkcolor=blue, citecolor=blue, urlcolor=blue}

  \makeatletter
    \renewcommand\@make@capt@title[2]{%
     \@ifx@empty\float@link{\@firstofone}{\expandafter\href\expandafter{\float@link}}%
      {\textsc{#1}}\@caption@fignum@sep#2\quad}%
    \makeatother

\newcommand{\KP}{K^\prime}
\newcommand{\kup}{K\uparrow}
\newcommand{\kdown}{K\downarrow}
\newcommand{\kpup}{K^\prime\uparrow}
\newcommand{\kpdown}{K^\prime\downarrow}

\begin{document}


\title{Machine learning unveils multiple Pauli blockades in the transport spectroscopy of bilayer graphene double-quantum dots}


\author{Anuranan Das \footref{contrib}}

\affiliation{Department of Electrical Engineering, Indian Institute of Technology Bombay, Powai, Mumbai-400076, India}
\author{Adil Khan \footref{contrib}}
\affiliation{Department of Electrical Engineering, Indian Institute of Technology Bombay, Powai, Mumbai-400076, India}
\author{Ankan Mukherjee \footref{contrib}}
\affiliation{Department of Physics, Indian Institute of Technology Bombay, Powai, Mumbai-400076, India}
\author{Bhaskaran Muralidharan \footref{contrib}}
\email{bm@ee.iitb.ac.in}
\affiliation{Department of Electrical Engineering, Indian Institute of Technology Bombay, Powai, Mumbai-400076, India}
\affiliation{Centre of Excellence in Quantum Information, Computation, Science and Technology, Indian Institute of Technology Bombay, Powai, Mumbai-400076, India}

\footnotetext[5]{All the authors contributed equally and are listed in alphabetical order of their last names\label{contrib}}


\date{\today}

\begin{abstract}
Recent breakthroughs in the transport spectroscopy of 2-D material quantum-dot platforms have engendered a fervent interest in spin-valley qubits. In this context, Pauli blockades in double quantum dot structures form an important basis for multi-qubit initialization and manipulation. Focusing on double quantum dot structures, and the experimental results, we first build theoretical models to capture the intricate interplay between externally fed gate voltages and the physical properties of the 2-D system in such an architecture, allowing us to effectively simulate Pauli blockades. Employing the master equations for transport and considering extrinsic factors such as electron-photon interactions, we thoroughly investigate all potential occurrences of Pauli blockades. Notably, our research reveals two remarkable phenomena: (i) the existence of multiple resonances within a bias triangle, and (ii) the occurrence of multiple Pauli blockades. Leveraging our model to train a machine learning algorithm, we successfully develop an automated method for real-time detection of multiple Pauli blockade regimes. Through numerical predictions and validations against test data, we identify where and how many Pauli blockades are likely to occur. We propose that our model can effectively detect the generic class of Pauli blockades in practical experimental setups and hence serves as the foundation for future experiments on qubits that utilize 2-D material platforms.
\end{abstract}

\maketitle


\section{Introduction\label{sec:introdution}}
\begin{figure*}[thpb]
\centering
\captionsetup[subfigure]{oneside,margin={0.3cm,-1cm}}
\captionsetup[subfloat]{oneside,margin={0.3cm,-1cm},labelfont=bf}
%
%
\begin{minipage}{\linewidth}
\centering
    \subfloat[]{
    \begin{minipage}{0.40\linewidth}
        \includegraphics[height=0.6\textwidth,width=\textwidth]{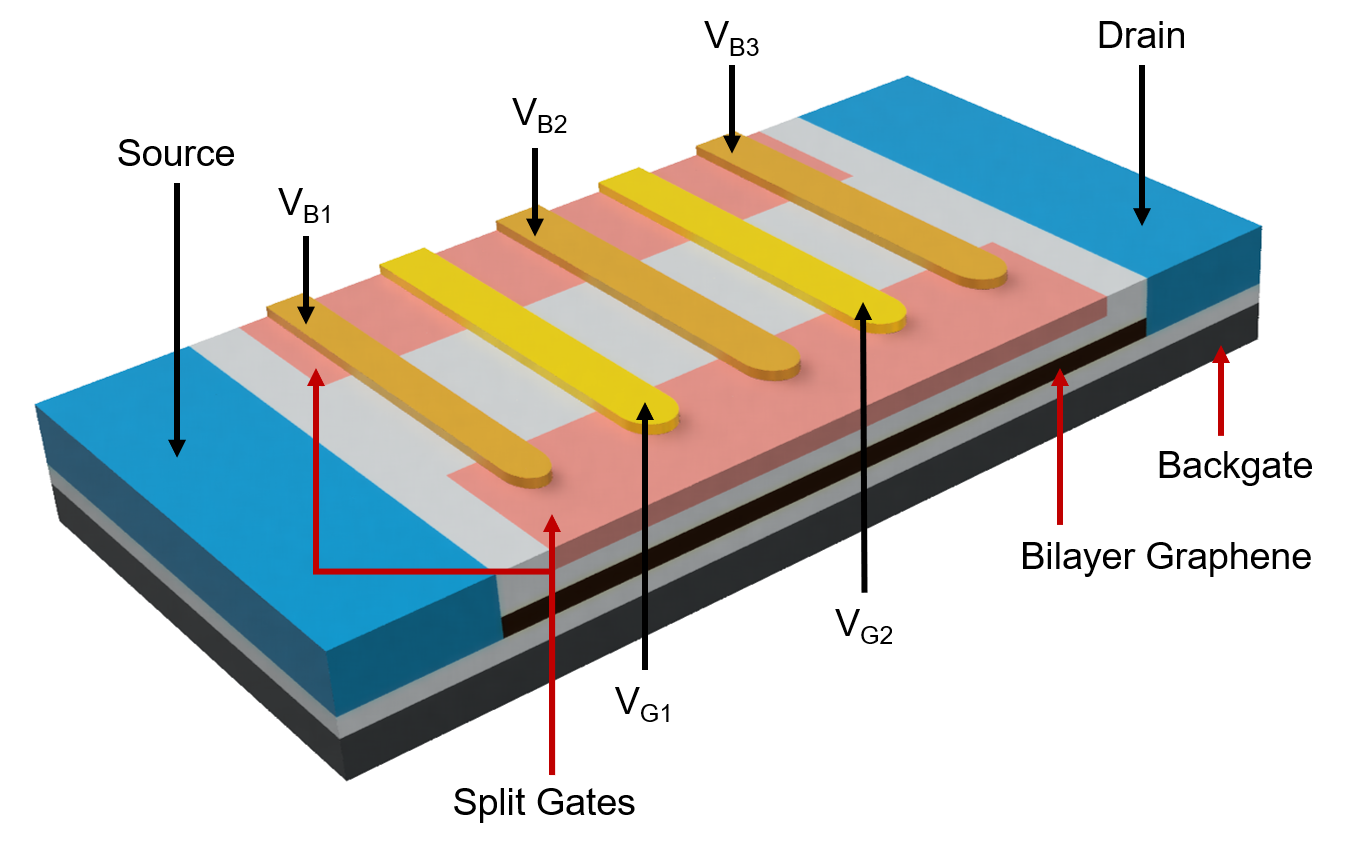}

    \end{minipage}
    \label{fig:3d_model}
    }
    \subfloat[]{
    \begin{minipage}{0.60\linewidth}
        \includegraphics[height=0.4\textwidth,width=0.8\textwidth]{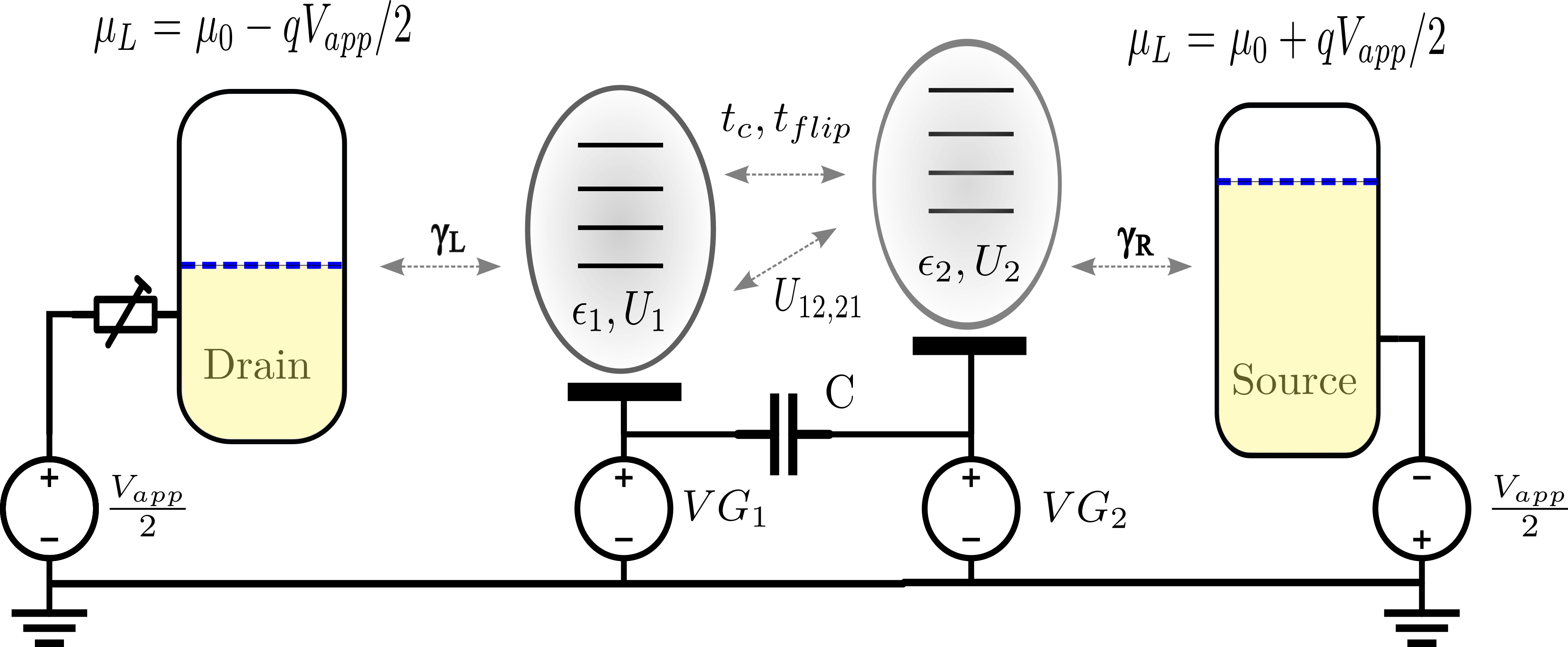}

    \end{minipage}\label{fig:2d model}}
    \\
    \subfloat[]{
    \begin{minipage}{0.4\linewidth}
        \includegraphics[height=0.55\textwidth, width=\textwidth]{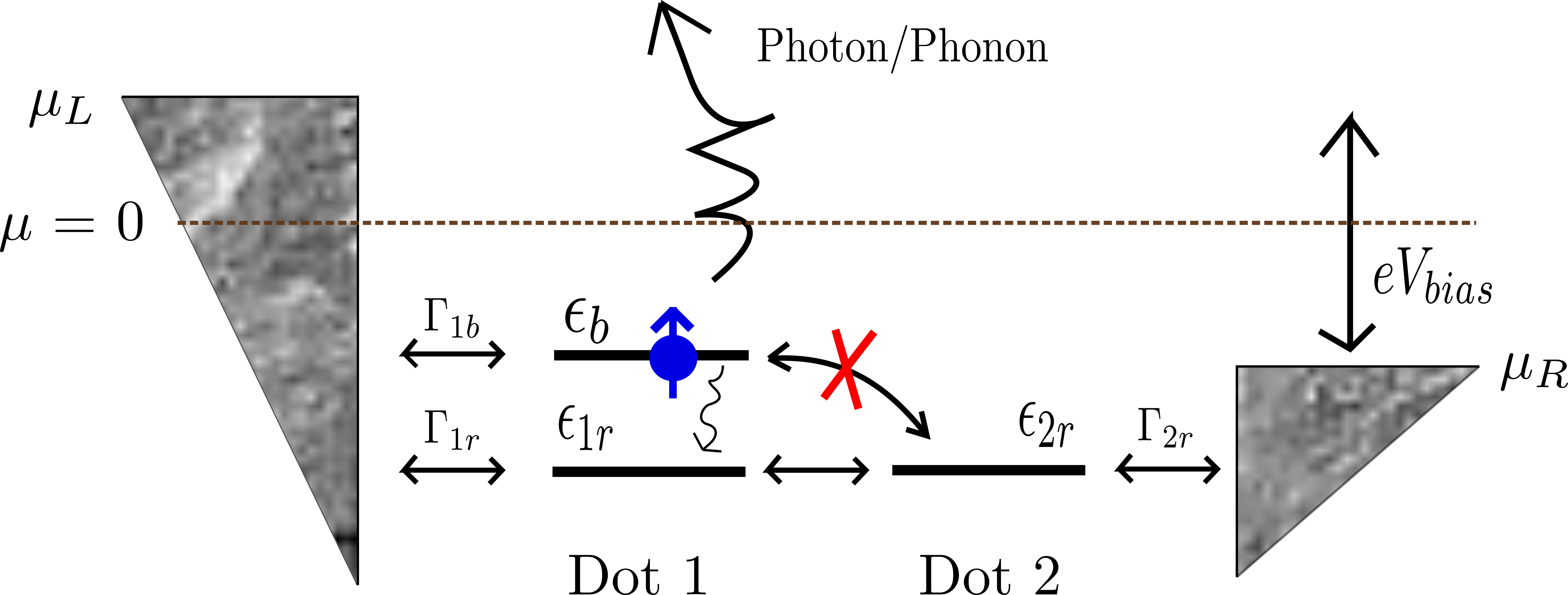}

    \end{minipage}\label{fig:PAT model}}
    \subfloat[]{
    \begin{minipage}{0.6\linewidth}
        \includegraphics[height=0.4\textwidth, width=0.8\textwidth]{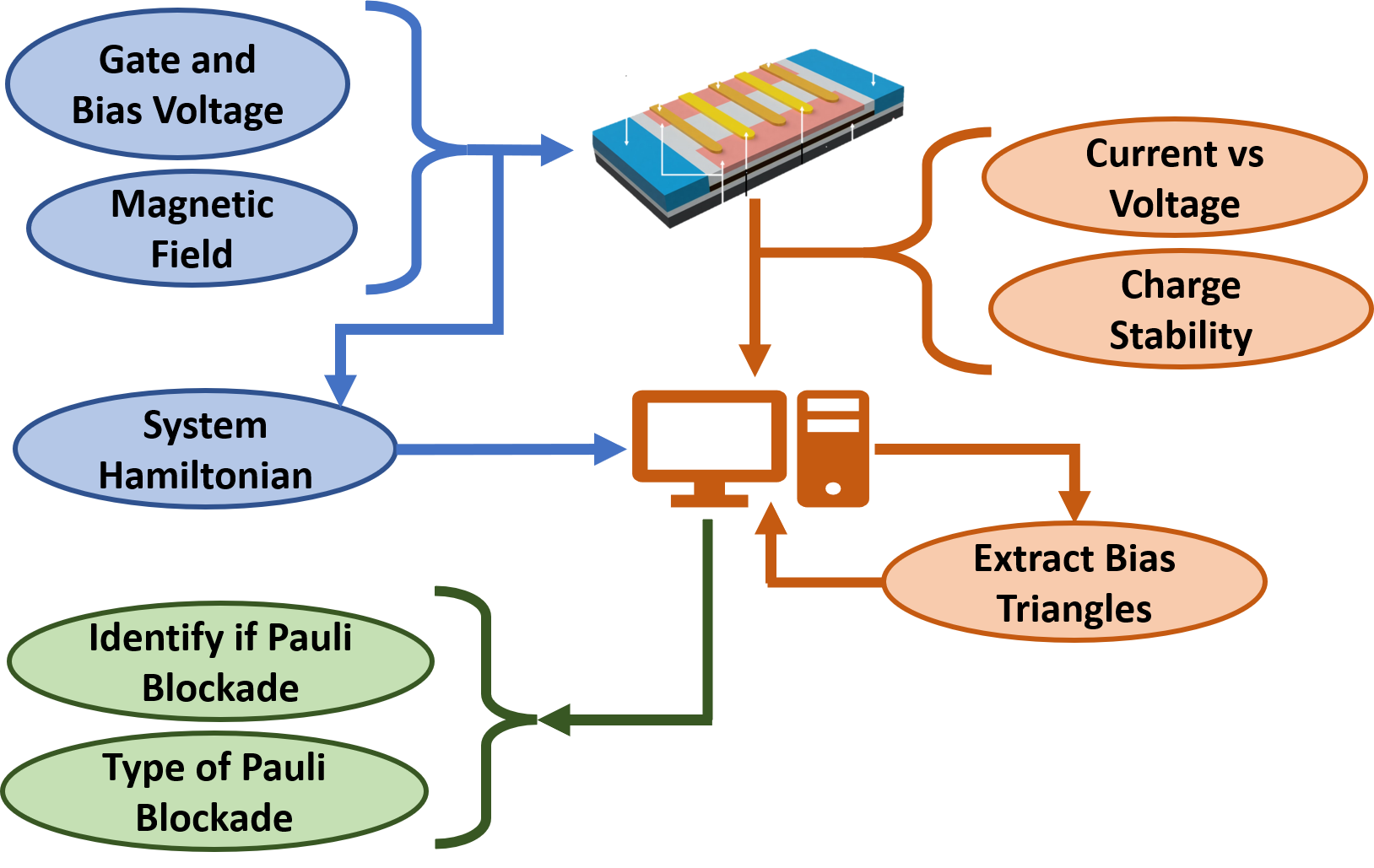}

    \end{minipage}\label{fig:flowchart}}
\end{minipage}
%
%
\captionsetup{justification=justified,singlelinecheck=false}
\caption{
Device schematics and workflow. 
\textbf{(a)} The device structure under consideration. The five finger gates and two split gates define the DQD transport spectroscopy setup. Barrier gates $V_{B1}$ and $V_{B3}$ control the coupling to the source and drain respectively, while $V_{B2}$ controls the inter-dot tunneling. Plunger gates $V_{G1}$ and $V_{G2}$ determine the onsite energies.
\textbf{(b)} The DQD system is coupled to the left and right contacts by coupling rates $\gamma_L$ and $\gamma_R$, respectively. Each dot is described by the onsite energies $\epsilon_i$, the intra-dot interaction $U_{ii}$, and the inter-dot interaction $U_{ij}$. The interdot tunneling is descrbed by $t_c$, which is spin and valley conserving. Spin-flip tunneling events are captured by $t_{flip}$. The dot tuning is controlled via gate voltages $V_{G1}$ and $V_{G2}$. A finite source-drain bias of $V_{sd}$ is also applied to facilitate current flow. The cross-capacitance between the gate voltages is $C$. 
\textbf{(c)} Inelastic processes are crucial to realize bias triangles in the high-bias regime. A basic mechanism for boson-assisted tunneling is shown. Intra-dot transitions between $\epsilon_b$ and $\epsilon_{2r}$ is blocked. An electron in level $\epsilon_b$ can loose energy via these interactions to transit to level $\epsilon_{1r}$ and hence open conduction channels involving states $\epsilon_{1r}$ and $\epsilon_{2r}$.
\textbf{(d)} Flowchart of the machine learning algorithm followed. The voltages and magnetic field inputs (blue) set to the system determine the parameters in the Hamiltonian, which, coupled with the I-V characteristics of the system (orange), are fed in to extract bias triangles and determine Pauli blockade regimes (green) using a machine learning algorithm.
}
\end{figure*}

Quantum devices implemented on 2-D material platforms have garnered significant attention due to their unique characteristics, including non-equivalent valleys, strong spin-orbit coupling, electronically tunable bandgaps, and reduced hyperfine interactions~\cite{liu,Giustino2020,2d_roadmap1,graphene_2D,spin_qubits_in_graphene,graphene_single_shot_readout,graphene_coherence,2d_hyperfine3}. This makes materials like bi-layer graphene (BLG) and the transition metal dichalcogenide family (TMDCs) \cite{Radha} promising platforms for spin-valley qubits. Recent experiments have explored the potential of utilizing the valley degree of freedom for qubit initialization, storage, and readout. Achieving confinement of electrons to quantum dots (QDs) on 2-D materials has, however, presented several hurdles, such as defects, high effective electron masses, and challenges in achieving precise gate voltage tunability using ohmic contacts \cite{Pisoni2018, Qiu2013, Jariwala2013}.  \\
\indent Initialization and read-out of qubits on any platform involves the ascertained occurrence of Pauli-blockades \cite{ono,si_tadokoro,main_exp_blg,Si_lai,Pauli_graphene,Stampfer_e_h_symmetry,tong2023} in the electronic transport spectroscopy. Recent investigations have demonstrated that in 2-D systems, the spin and the valley pseudospin are intertwined, leading to a general class of Pauli blockades ~\cite{Mukherjee2023} that cannot be solely explained by independently considering the spin or valley degrees of freedom. While the theoretical understanding of the origins of multiple Pauli blockades has made progress~\cite{garreis2023, tong2023}, leveraging this knowledge to tune quantum dots by identifying and characterizing multiple blockades still presents a significant challenge. \\
\indent Utilizing machine learning for the initialization and scaling of qubits in quantum devices have highlighted the use of automation to detect and characterize Pauli spin-blockades in semiconducting qubit systems~\cite{Ares2021, Ares2023, Gebhart2023, schuff2022}. These algorithms can swiftly determine the occurrence of a spin-blockade, enabling real-time readout of spin qubits. 
Given the rapid progress in spin-valley qubit research, a fully automated algorithm for identifying multiple spin-valley coupled Pauli blockades is certainly the need of the hour. In this paper, we develop a comprehensive model to understand the transport mechanism through quantum dots. We leverage this model to train a machine learning algorithm, enabling us to predict blockade regimes in the 2-D material platforms, as illustrated in Fig.~\ref{fig:flowchart}. \\
\indent Our work builds upon a double quantum dot (DQD) transport setup \cite{dqd_structure1, dqd_structure2, BM_1, BM_2007,Muralidharan_2008}, as depicted in Fig.~\ref{fig:3d_model}. The focus of our analysis is on Pauli blockades, particularly in the regime where the total occupancy of the dots is restricted to two electrons. To construct our models, we account for various factors~\cite{hubbard1,hubbard2,hubbard3,TMDC_paper} such as intrinsic spin-orbit (SO) coupling, spin-Zeeman splitting, valley-Zeeman splitting, sequential spin (pseudo-spin) conserving inter-dot tunneling, spin-flip tunneling, photon-assisted tunneling, and Coulomb interactions. These elements are visualized in Fig.~\ref{fig:2d model} and Fig.~\ref{fig:PAT model}, and are the essential components of our model for understanding the intricacies of the transport mechanism, and eventually making important predictions.

By simulating the effects of spin-orbit coupling, spin-flip tunneling and photon-assisted tunneling, we unveil two novel phenomena, which are not present in spin-only systems: (i) multiple resonances and (ii) multiple Pauli blockades. We demonstrate that these phenomena have distinct underlying causes. Furthermore, we carefully train our machine learning model to not only detect the presence of a Pauli blockade but also predict whether there are multiple Pauli blockades or just multiple resonance peaks in the system. The predictability of precise Pauli blockade regimes allows us to extract both spin and valley information from the state using a single-shot readout, effectively doubling the rate at which information is stored and read. We firmly believe that our research lays the groundwork for future experiments involving the initialisation and scaling of qubits on 2-D material platforms.\\
\indent The paper is organised as follows. In Sec.~\ref{sec:formalism}, we construct the effective Hamiltonian for the DQD, solve for the relevant eigenstates, and obtain the equations governing the current flow and thereby obtain the transport spectroscopy. We then present the conditions under which Pauli blockades may be realized, and under which they may be lifted. In Sec.~\ref{sec:results} we perform simulations on our model and generate a training data set using different regimes and parameters. We also present the cases under which multiple excitations and multiple blockades occur. We then use the generated data to train our machine learning model and then test the same on different test cases. Finally, we present our conclusions and scope for future work in Sec.~\ref{sec:conclusion}.
\section{Formalism and Methods\label{sec:formalism}}
A schematic of the model under investigation is illustrated in 
Fig.~\ref{fig:3d_model}, and Fig.~\ref{fig:2d model}. The model involves a DQD setup~\cite{DQD1,DQD2} with source and drain electrodes shown in blue. Split gates (pink) are used to change the carrier concentration, and finger gates (yellow) are used for fine-tuning the onsite potentials on each quantum dot. Functionally, when a dc bias is applied across this device, a current can be observed. The current flow can then also be controlled by tweaking the gate voltages. We have a left (right) quantum dot $LD (RD)$, whose onsite energy, $\varepsilon_1 (\varepsilon_2)$ can be controlled using the lever arm voltage $V_{G1}(V_{G2})$. The model abstraction for the DQD system is illustrated in Fig.~\ref{fig:3d_model}. The source drain bias voltage $V_{SD}$ is applied across the electrodes labelled $V_S$ and $V_D$. The voltage at gate $V_{B2}$ controls $t$, the inter-dot tunneling, which, in our model, preserves the spin and the valley pseudo-spin. The coupling constants between each lead and the channel, $\gamma_1$ and $\gamma_2$, are controlled by barrier gate voltages $V_{B1}$ and $V_{B3}$. The backgate voltage $V_{BB}$ can be tuned to control the overall degree of confinement.
\subsection{Model Hamiltonian\label{subsec:formalism:model}}
We employ a modified version of the well-studied Hubbard Hamiltonian~\cite{qtt, TMDC_paper, Mukherjee2023} to model the system. This model includes several components, such as onsite energies and inter-dot tunneling. Additionally, there exists an onsite Coulomb repulsion, denoted by $U_{1(2)}$, between pairs of electrons on each dot, and an inter-dot Coulomb repulsion, denoted by $U_{12(21)}$. These energies $U_i$ and $U_{ij}$ can be matrices indexed by $i(=1,2)$ and $j(=1,2)$, allowing for a more flexible representation. \\
\indent Each dot can host a conduction electron in one of the two valleys, either $K$ or $\KP$, and each valley can possess either spin $\uparrow$ or spin $\downarrow$. Consequently, a single dot offers four available states: $\ket{\kup}$, $\ket{\kdown}$, $\ket{\kpup}$, and $\ket{\kpdown}$. In the absence of an external magnetic field, these four energy states split into two Kramer pairs due to intrinsic spin-orbit (SO) coupling~\cite{so_theory_blg1, so_theory_blg2, so_exp_blg, so_exp1, so_exp2}. However, when an external magnetic field is introduced, the degeneracy of the states within each Kramer pair is broken, affected by both the spin-Zeeman and the valley-Zeeman effects. The energy shift due to spin-Zeeman splitting is given by $h_S = \sigma g_s \mu_B B_{||}$, while the valley-Zeeman splitting contributes an energy shift of $h_V = \tau g_v \mu_B B_\perp$. Here, $\sigma$ represents the spin quantum number ($\sigma=+\frac{1}{2}$ for spin $\uparrow$ and $\sigma=-\frac{1}{2}$ for spin $\downarrow$), and $\tau$ denotes the valley pseudo-spin ($\tau=+\frac{1}{2}$ for valley $K$ and $\tau=-\frac{1}{2}$ for valley $\KP$). The electron magnetic moment is denoted by $\mu_B=5.79\times10^{-5}$ eVT$^{-1}$, $B_{||}$ and $B_\perp$ are the parallel and perpendicular components of the external magnetic field, and $g_s$ and $g_v$ represent the spin and valley g-factors, respectively. 


Under such considerations, the Hamiltonian takes the form
\begingroup
\allowdisplaybreaks
\begin{align}
    \hat{H}_{DQD} &= \underbrace{\varepsilon_L \hat{n}_L + \varepsilon_R \hat{n}_R}_{\text{Onsite energy}}\nonumber\\
    &\hspace{2mm}+\underbrace{\frac{U_{ii}}{2}\left(\hat{n}_L^2-\hat{n}_L+\hat{n}_R^2-\hat{n}_R\right)}_{\text{Onsite repulsion}}%
    +\underbrace{U_{ij}\hat{n}_L\hat{n}_R}_{\text{Interdot replusion}}\nonumber\\
    &\hspace{4mm}+\underbrace{\sum_{\tau,\sigma}t\hat{c}_{R\sigma\tau}^\dagger \hat{c}_{L\sigma\tau} + \text{h.c.}}_{\text{Interdot sequential tunneling}}\nonumber\\
    &\hspace{6mm}+\underbrace{\sum_{\alpha,\tau,}t_{SO}\hat{c}_{R\uparrow\tau}^\dagger \hat{c}_{L\downarrow\tau} + \hat{c}_{R\downarrow\tau}^\dagger \hat{c}_{L\uparrow\tau} + \text{h.c.}}_{\text{Interdot spin-flip tunneling}}\nonumber\\
    &\hspace{8mm}+\underbrace{\frac{\Delta_{SO}}{2}\sum_{\alpha,\tau,\sigma}\hat{c}_{\alpha\sigma\tau}^\dagger\left(\bm{\sigma}_3\right)_{\sigma\sigma}\left(\bm{\tau}_3\right)_{\tau\tau}\hat{c}_{\alpha\sigma\tau}}_{\text{Spin orbit coupling}}\nonumber\\
    &\hspace{10mm}+\underbrace{\left|h_S\right|\sum_{\alpha,\tau,\sigma}\hat{c}_{\alpha\sigma\tau}^\dagger\left(\bm{\sigma}_3\right)_{\sigma\sigma}\hat{c}_{\alpha\sigma\tau}}_{\text{Spin Zeeman effect}}\nonumber\\
    &\hspace{10mm}+\underbrace{\left|h_V\right|\sum_{\alpha,\tau,\sigma}\hat{c}_{\alpha\sigma\tau}^\dagger\left(\bm{\tau}_3\right)_{\tau\tau}\hat{c}_{\alpha\sigma\tau}}_{\text{Valley Zeeman effect}}\label{eq:fermi_hubbard_hamiltonian},
\end{align}%
\endgroup
where the summations are defined over $\alpha\in\{L,R\}$, $\sigma\in\{\uparrow,\downarrow\}$, $\tau\in\{K,\KP\}$. The terms $\bm{\sigma}_3$ ($\bm{\tau}_3$) is the z-component of the Pauli matrix for the spin (valley pseudo-spin), defined as 
\begin{subequations}
\begin{alignat}{4}
    \left(\bm{\sigma}_3\right)_{\uparrow\uparrow}&=1;\quad\left(\bm{\sigma}_3\right)_{\uparrow\downarrow}&&=\left(\bm{\sigma}_3\right)_{\downarrow\uparrow}&&=0;\quad\left(\bm{\sigma}_3\right)_{\downarrow\downarrow}&&=-1\\
    \left(\bm{\tau}_3\right)_{KK}&=1;\quad\left(\bm{\tau}_3\right)_{K\KP}&&=\left(\bm{\tau}_3\right)_{\KP K}&&=0;\quad\left(\bm{\tau}_3\right)_{\KP\KP}&&=-1
\end{alignat}
\end{subequations}
The symbol $\hat{n}_L$ ($\hat{n}_R$) denotes the number operator for the number of electrons on the left (right) dot, formulated as
\begin{align}
    \hat{n}_\alpha=\sum_{\sigma,\tau}\hat{c}^\dagger_{\alpha\sigma\tau}\hat{c}_{\alpha\sigma\tau},
\end{align}
where $\hat{c}^{(\dagger)}_{\alpha\sigma\tau}$ is the annihilation (creation) operator for an electron on dot $\alpha$ with spin $\sigma$ in valley $\tau$.

In the absence of spin-flip tunneling and photon interactions, the above Hamiltonian takes a block diagonal form and separates into nine sub-spaces, each with an invariant total number of electrons $N=0,1,\cdots,8$. In discussing the blockades in the two-electron occupancy regime, only the subspaces $N=1$ and $N=2$ are relevant.
We use $L_\zeta\left(R_\zeta\right)$ to denote that there is an electron in the state $\zeta$ on the left(right) quantum dot, where $\zeta\in\{\kup, \kdown, \kpup, \kpdown\}$. For instance, a system with two electrons: one in the left dot in state $\kup$ and the other in the right dot in state $\kpdown$ is represented as $\ket{L_{\kup} R_{\kpdown}}$. We also develop the notation $(n_L,n_R)$ to represent a state with $n_L$ electrons on the left dot and $n_R$ electrons on the right dot. The eigenstates are not the $(n_L,n_R)$ states, but a superposition of $(n_L,n_R)$ states with $n_L+n_R=N=\text{constant}$.

\begin{figure*}[thpb]
\centering
\captionsetup[subfigure]{oneside,margin={0.3cm,-1cm}}
\captionsetup[subfloat]{oneside,margin={0.3cm,-1cm},labelfont=bf}
%
%
\centering
    \subfloat[]{
    \begin{minipage}{0.40\linewidth}
        \includegraphics[width=\textwidth]{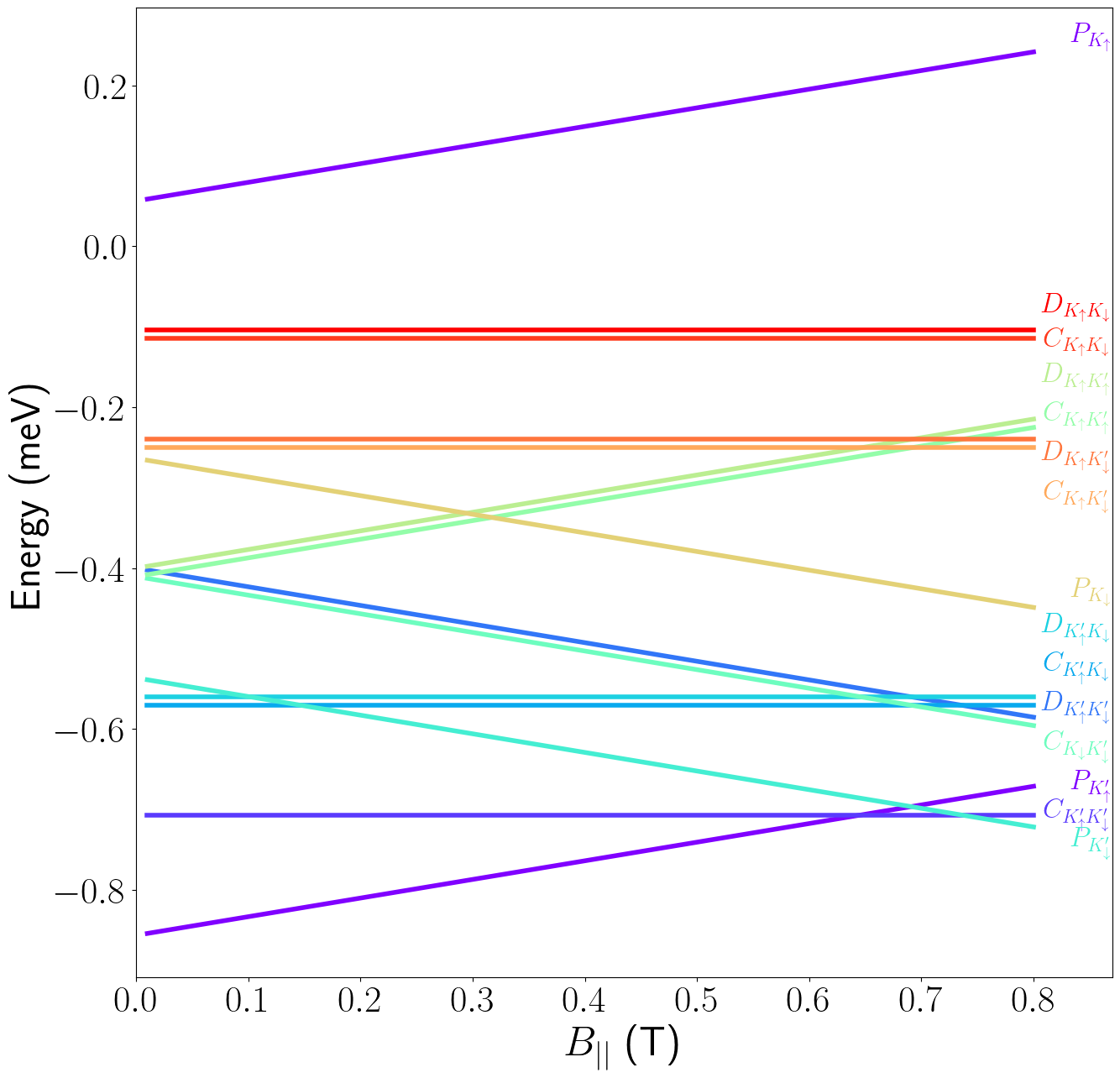}

    \end{minipage}
    \label{fig:evsb_par}
    }
    \hspace{2cm}
    \subfloat[]{
    \begin{minipage}{0.40\linewidth}
        \includegraphics[width=\textwidth]{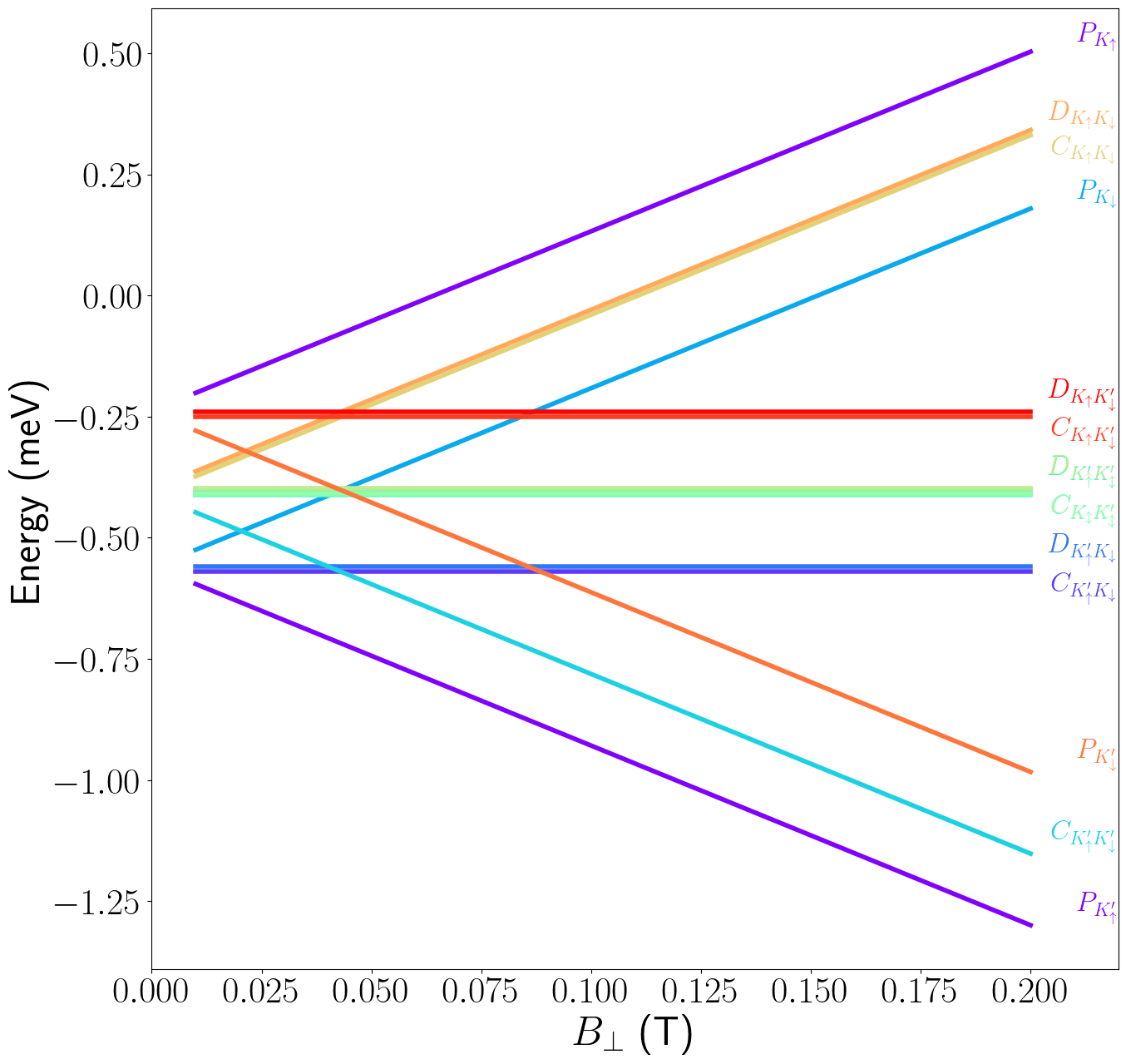}
    \end{minipage}\label{fig:evsb_perp}}
\captionsetup{justification=justified,singlelinecheck=false}
\caption{
Energy spectrum of the $N=2$ Fock space as is described in Sec.~\ref{subsec:formalism:fock}, as a function of the magnetic field applied.
\textbf{(a)} The perpendicular component of the magnetic field is held fixed at $B_\perp = 0.8$ T, and the parallel component is varied. The spin Zeeman effect splits the energy spectrum according to $\Delta E=\mu_B g_s B_{||}$.
\textbf{(b)} The parallel component of the magnetic field is held fixed at $B_{||} = 0.8$ mT, and the perpendicular component is varied. The valley Zeeman effect splits the energy spectrum according to $\Delta E=\mu_B g_v B_\perp$.
}
\end{figure*}
In addition to the above Hamiltonian, we also add a term for photon-assisted tunneling (PAT) as described below. These additional effects often lift the Pauli blockades by creating eigenstates that are superpositions of multiple spin and valley states. 
The composite Hamiltonian of the system can be modified to include the effects of boson-mediated scattering processes~\cite{Aghassi2007ElectronicTA} as shown in fig.~\ref{fig:PAT model}. The photon relaxation occurs in the sense that electrons in the dot can change level by absorption and emission. Such an interaction Hamiltonian is described as:
\begin{equation}
H_{\mathrm{ph}}=H_B+H_{B-D}=\sum_q \omega_q d_q^{\dagger} d_q+\sum_{q \sigma i j} g_{p h}\left(d_q^{\dagger}+d_q\right) c_{i \sigma}^{\dagger} c_{j \sigma},
\end{equation}
where the coupling amplitudes are independent of quantum numbers i, j and q. Hence, $g_q^{i j}=g_{p h}$. Total bosonic coupling constant is thus $\alpha_{\mathrm{ph}}(\omega)=2 \pi g_{\mathrm{ph}}^2 \rho_{\mathrm{b}}(\omega)$, where $\rho_{\mathrm{b}}(\omega)$ represents the photon density of states. 

\begin{figure*}[thpb]

\captionsetup[subfigure]{oneside,margin={0.3cm,-1cm}}
\captionsetup[subfloat]{oneside,margin={0.3cm,-1cm},labelfont=bf}
%

\begin{minipage}{0.32\linewidth}
\centering
\subfloat[]{\includegraphics[width=\linewidth]{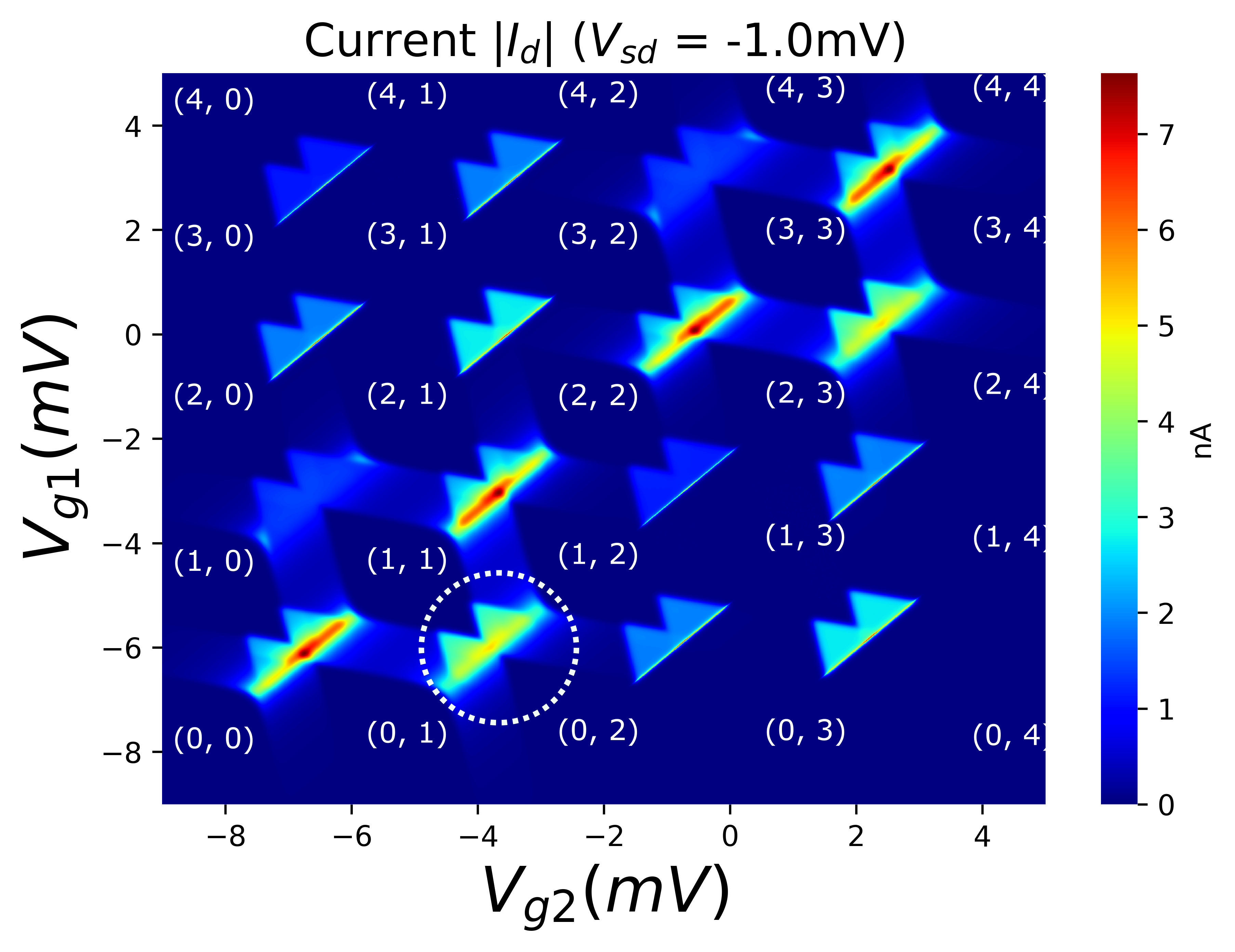}
    \label{fig:high_b_cias}}\\
\subfloat[]{\includegraphics[width=\linewidth]{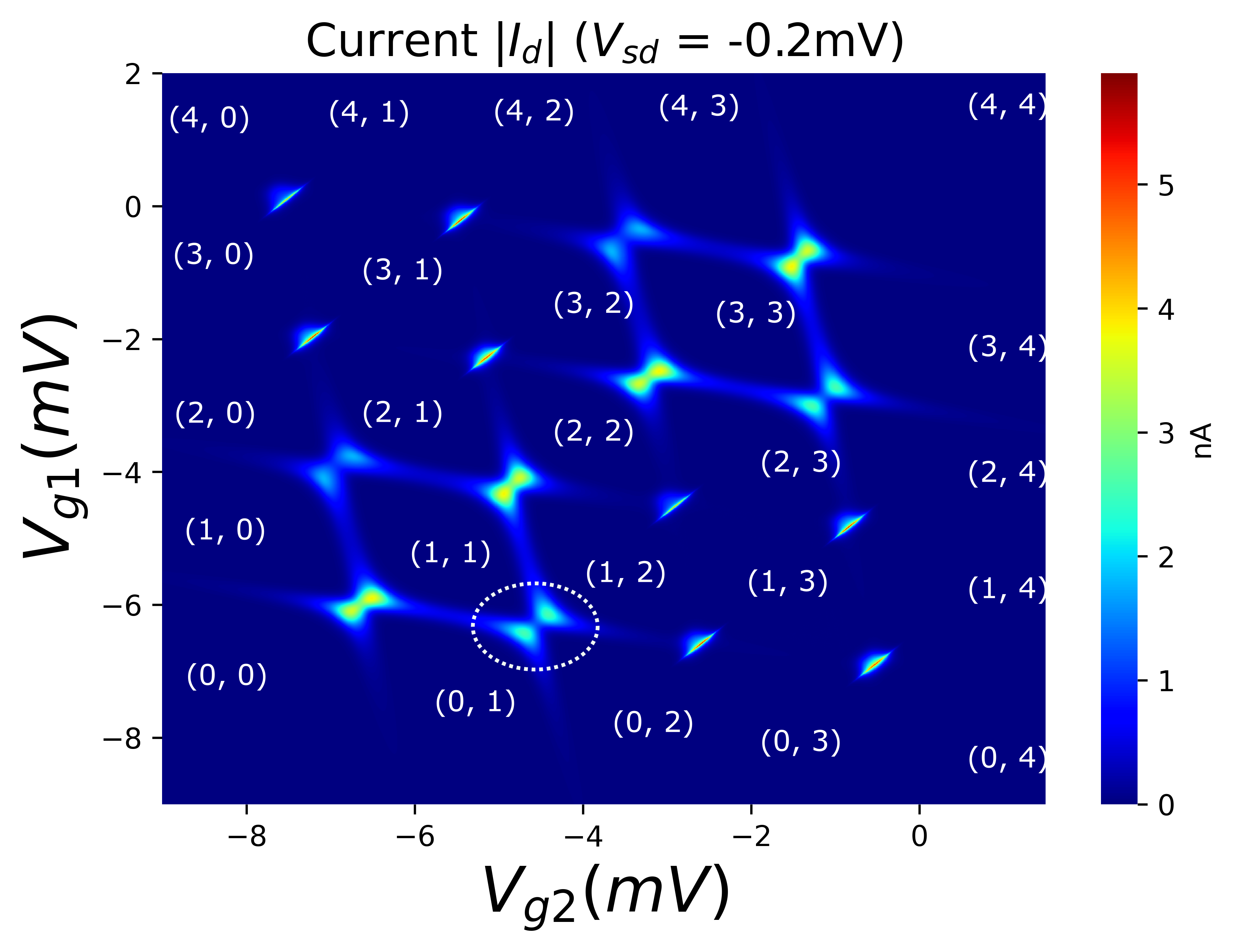}
    \label{fig:low_b_cias}}
\end{minipage}
\begin{minipage}{0.32\linewidth}
\centering
    \subfloat[]{\includegraphics[width=\linewidth]{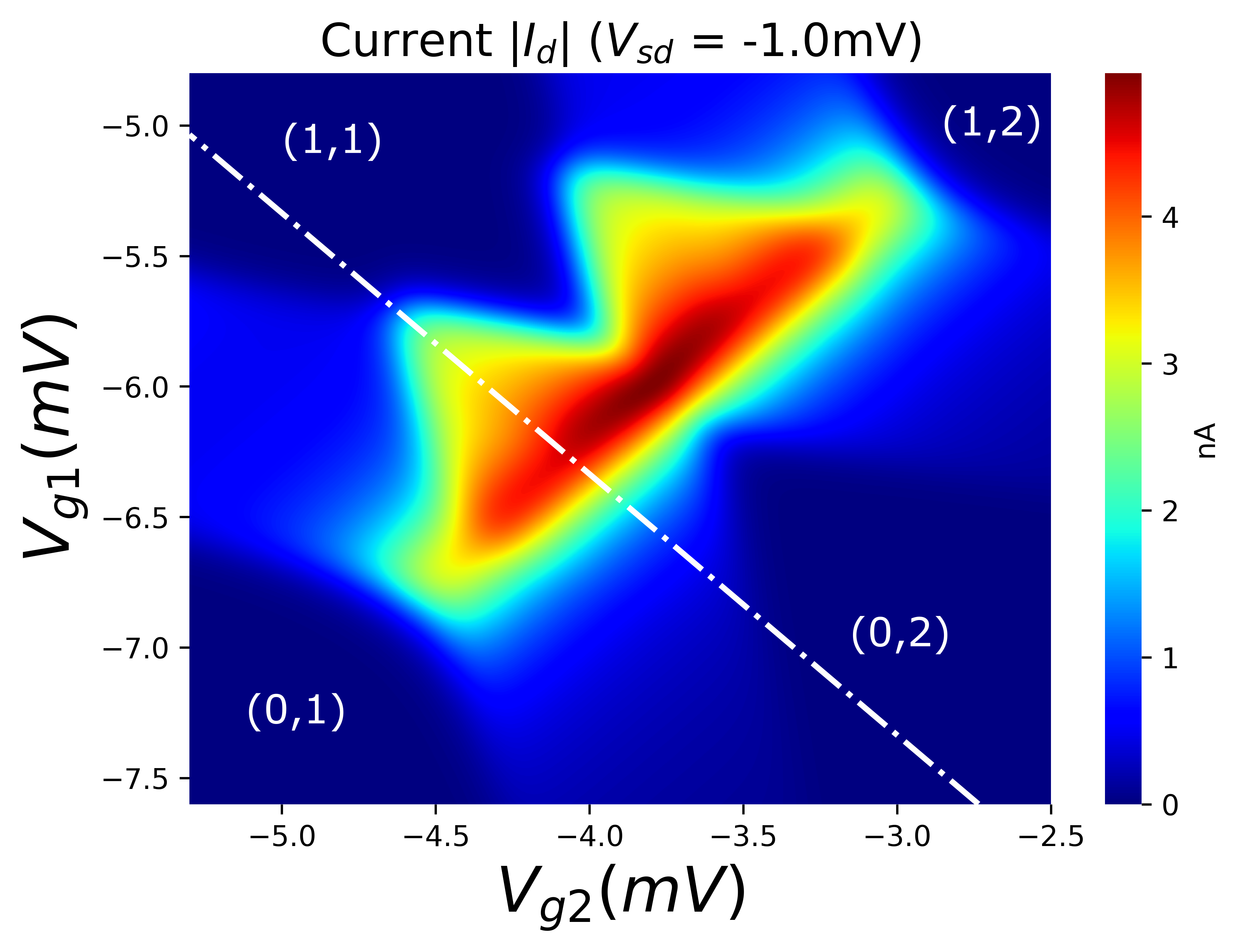}
    \label{fig:for_pat}}\\
    \subfloat[]{\includegraphics[width=\linewidth]{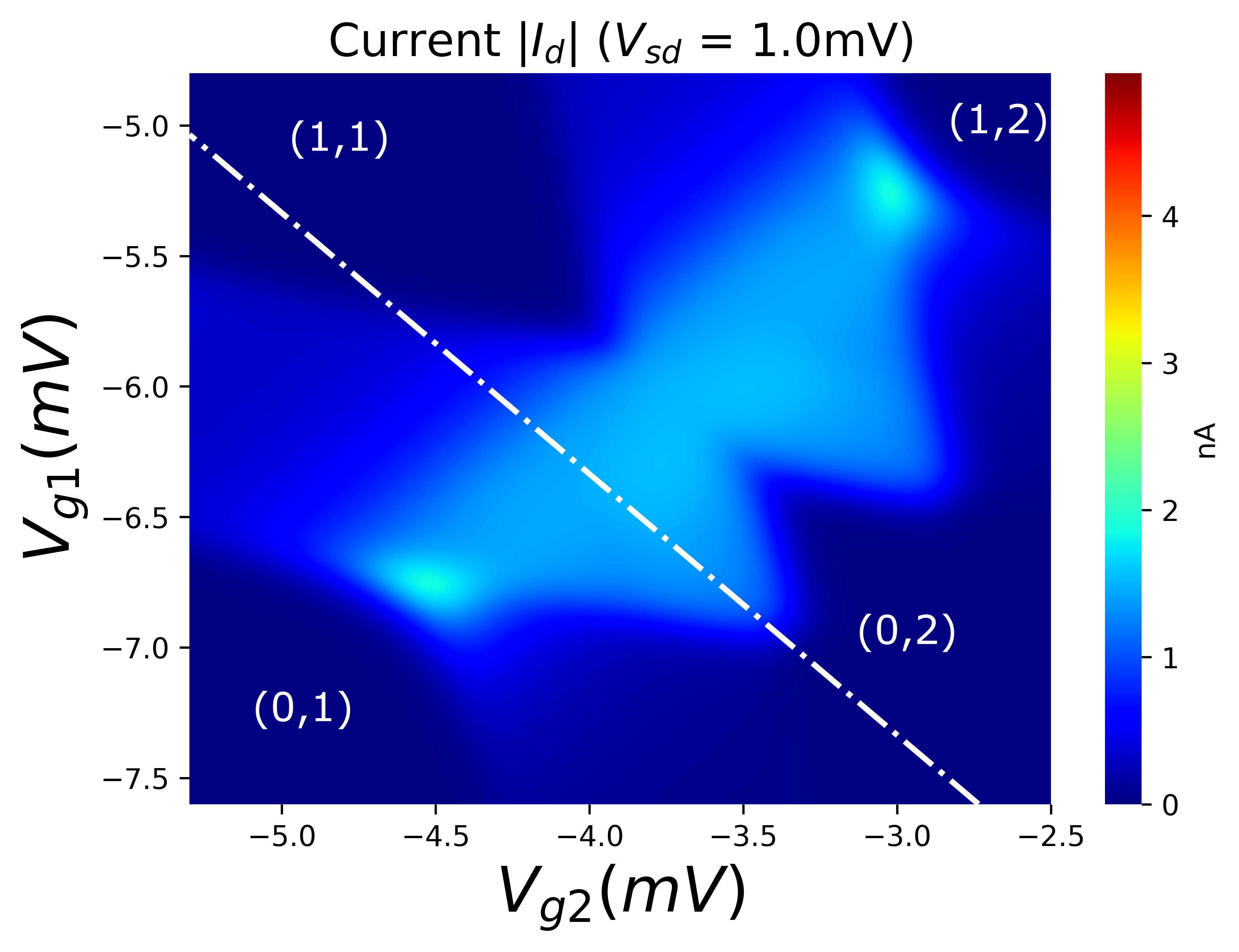}
    \label{fig:rev_pat}}
\end{minipage}
\begin{minipage}{0.32\linewidth}
\centering
    \subfloat[]{\includegraphics[width=\linewidth]{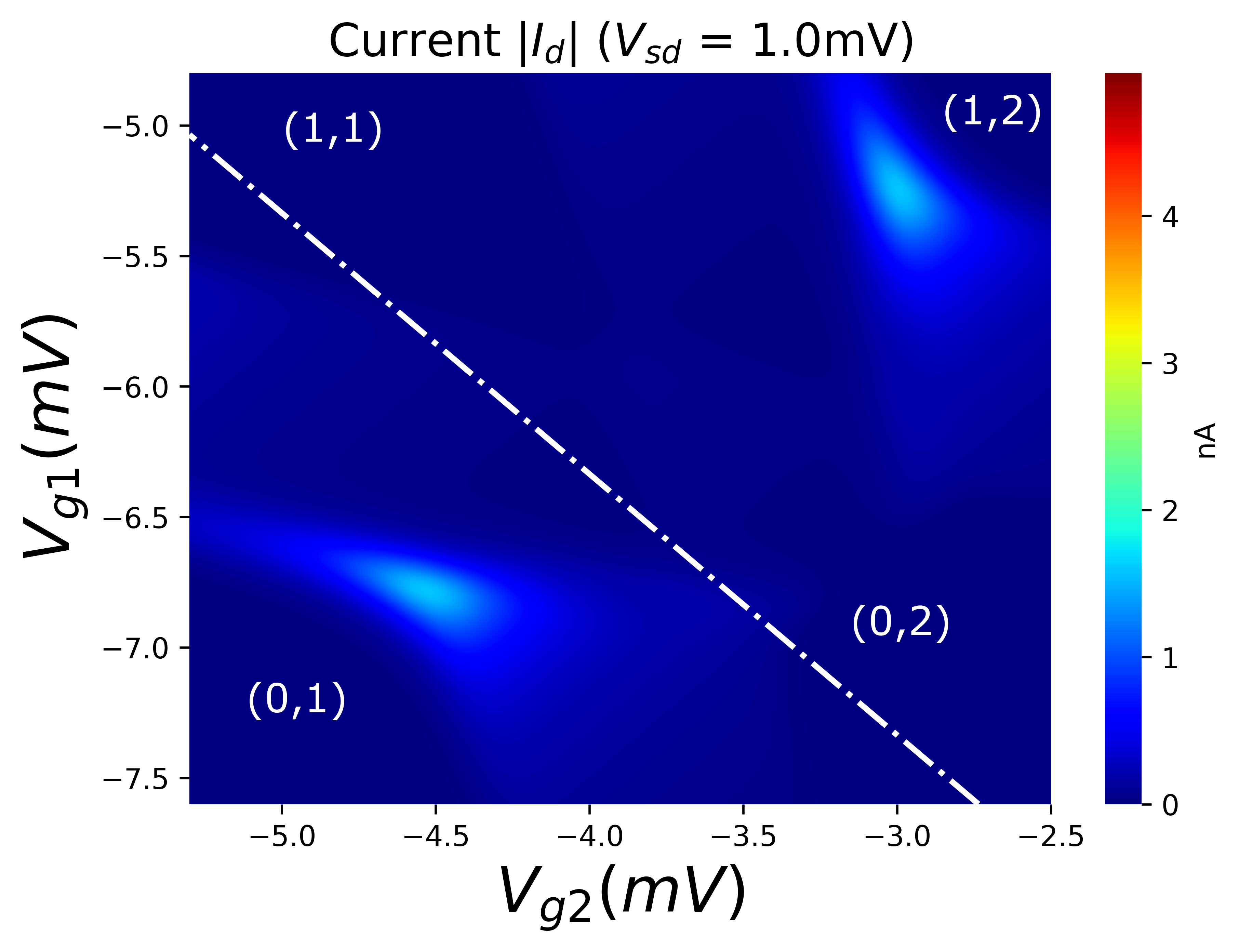}
    \label{fig:rev_nopat}}\\
    \subfloat[]{
    \includegraphics[width=\linewidth]{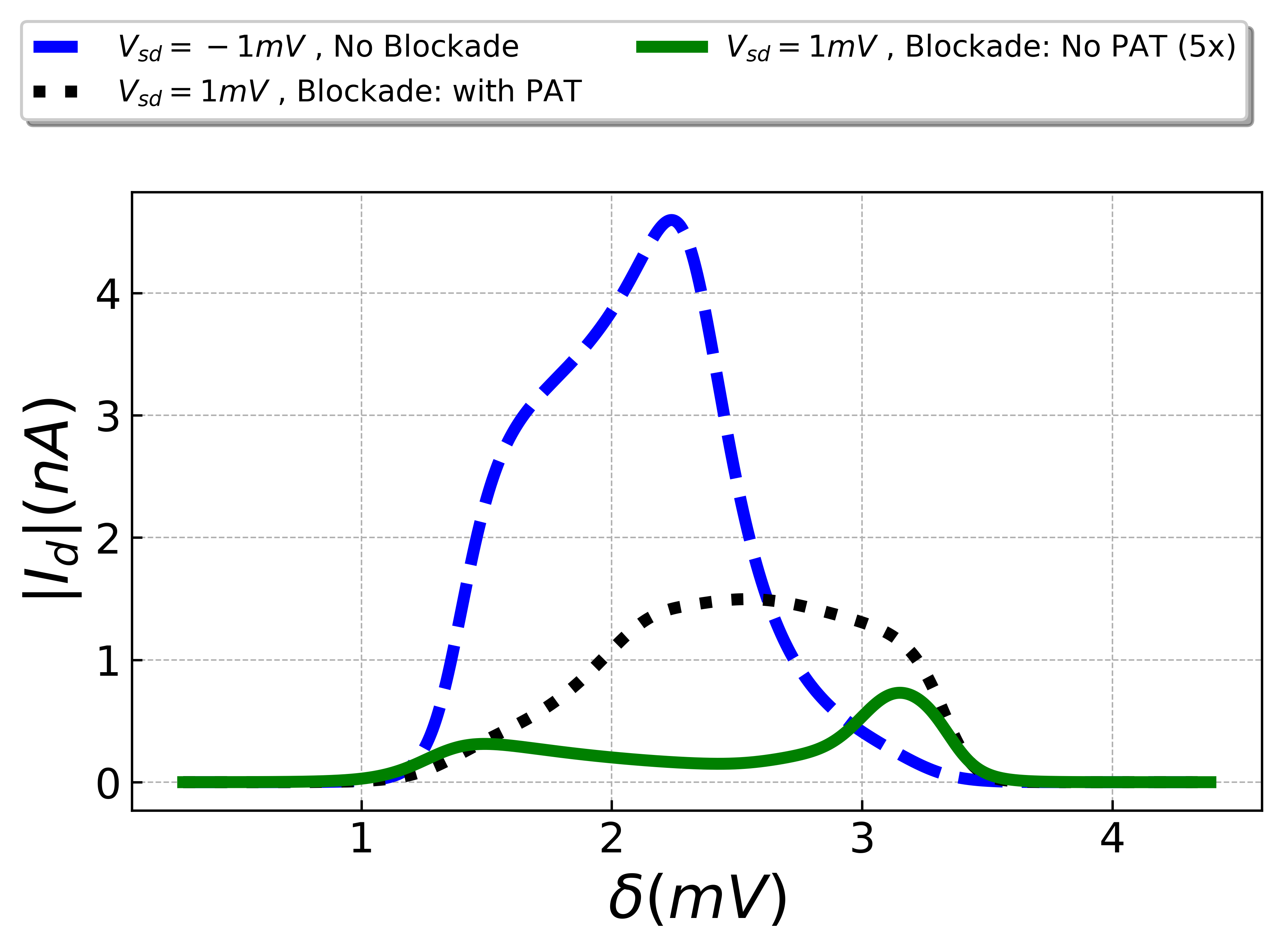}
    \label{fig:detunes}
    }
\end{minipage}
%

%
\captionsetup{justification=raggedright,singlelinecheck=false}
\caption{
Analysis of charge stability(CS) diagrams with emphasis on the (1,1) to (0,2) transition and the effect of various physical parameter assumptions.
\textbf{(a)} CS diagrams with clearly defined bias triangles can be obtained in the high source-drain bias regime($V_{sd} = -1.0 mV$). 
\textbf{(b)} At relatively lower biases($V_{sd} = -0.2 mV$) however, the triangles are not formed. For both the CS diagrams,  the electron occupancy in the dots is marked.
\textbf{(c)} Bias triangles magnified from the encircled part in (a) for (1,1) to (0,2) transition. For $V_{sd} = -1 mV$ there is no blockade. 
\textbf{(d)} Reversing the bias($V_{sd} = -1 mV$) leads to blockade and a four-fold reduction in current is seen.
\textbf{(e)} On turning off PAT, further reduction in current(by 10 times) is noticed.
\textbf{(f)} Current vs detuning plot obtained by taking cross-section along the white lines in figures (c), (d), and (e).
The blockade mechanism is explained in Sec.~\ref{subsec:results: blockade}.
}
\label{fig:Origin_BT}
\end{figure*}

\subsection{Fock subspaces of the Hamiltonian\label{subsec:formalism:fock}}
In the presence of sequential tunneling alone, the Hamiltonian is a block diagonal matrix, each block representing the total number of electrons, $N$ in the DQD. Exact solutions to the eigenstates have emerged recently~\cite{Mukherjee2023}.
The $N=1$ sub-matrix of the Hamiltonian in ~\eqref{eq:fermi_hubbard_hamiltonian} is an $8\times8$ matrix and consequently has eight distinct eigenstates. These eigenstates can be categorized into two groups: the bonding states and the anti-bonding states, as follows.
\begin{subequations}
\begin{alignat}{2}
\ket{B_{\zeta}} &= \xi\ket{L_\zeta}+\eta\ket{R_\zeta}\\
\ket{AB_{\zeta}} &= \xi\ket{L_\zeta}-\eta\ket{R_\zeta}.
\end{alignat}
\label{eq:N_1_fockspace}
\end{subequations}
In the above formulation, $B$ represents the bonding states and are lower in energy, while $AB$ represents the anti-bonding states and are higher in energy. $\zeta\in\{\kup, \kdown, \kpup, \kpdown\}$.  
\indent The $N=2$ sub-matrix of the Hamiltonian is a $28\times28$ matrix with twenty eight eigenstates. Based on their contribution towards the current through the DQD, the eigenstates can be classified into three broad categories as follows.
\begin{subequations}
\begin{alignat}{2}
\ket{C_{\zeta_1\zeta_2}} &= \alpha\left(\ket{L_{\zeta_1}R_{\zeta_2}}-\ket{L_{\zeta_2}R_{\zeta_1}}\right)\nonumber\\
&\hspace{20mm}+\beta\ket{L_{\zeta_1}L_{\zeta_2}}+\kappa\ket{R_{\zeta_1}R_{\zeta_2}}\\
\ket{D_{\zeta_1\zeta_2}} &= \frac{1}{\sqrt{2}}\left(\ket{L_{\zeta_1}R_{\zeta_2}}+\ket{L_{\zeta_2}R_{\zeta_1}}\right)\\
\ket{P_{\zeta}} &= \ket{L_{\zeta}R_{\zeta}},
\end{alignat}
\label{eq:N_2_fockspace}
\end{subequations}
where $\zeta\in\{\kup, \kdown, \kpup, \kpdown\}$ and the combination $\zeta_1\zeta_2\in\{\kup\kdown,\ \kup\kpup,\ \kup\kpdown,\ \kdown\kpup,\ \kdown\kpdown,\ \kpup\kpdown\}$ and $\zeta_1$ and $\zeta_2$ are taken over all possible unordered combinations of $\zeta$. We therefore have six possible combinations of $\zeta_1\zeta_2$. Each of the states $C$ occurs threefold, with three different sets of values of $\alpha$, $\beta$, and $\kappa$. Thus, in total, we have eighteen $C$ states, six $D$ states, and four $P$ states. The state labels $C$ and $D$ stand for "conducting" and "dark" respectively, according to their role in the equation for the current~\cite{Mukherjee2023}. \\
\indent In the presence of spin-flip tunneling, the states belonging to the $N=1$ Fock-space are superpositions of two bonding (or anti-bonding) states. The bonding superposition takes the form 
\begin{align}
    a\ket{B_{\zeta_1}}+b\ket{B_{\zeta_1}} &= \xi a\ket{L_{\zeta_1}}+\eta a\ket{R_{\zeta_1}}\nonumber\\
    &\hspace{6mm}+\xi b\ket{L_{\zeta_2}}+\eta b\ket{R_{\zeta_2}}\nonumber
\end{align}
and the antibonding state follows accordingly. In the $N=2$ subspace, we obtain superpositions of two $C$ or $D$ states in a similar fashion. The eigenvalue variation of various states in the $N=2$ manifold with respect to $B_{\parallel}$ and $B_{\perp}$ is shown in Figs.~\ref{fig:evsb_par}(a) and (b) respectively. The existence of superposition increases the number of states that satisfy the transport selection rules \cite{BM_2007,Muralidharan_2008,Mukherjee2023} and provides multiple transition pathways to the same state, each having a different transition energy. This manifests as multiple resonance peaks within a bias triangle, each corresponding to a particular transition.

\begin{figure*}[thpb]

\captionsetup[subfigure]{oneside,margin={0.3cm,-1cm}}
\captionsetup[subfloat]{oneside,margin={0.3cm,-1cm},labelfont=bf}
%

\subfloat[]{
\begin{minipage}{0.32\linewidth}
\centering
    \includegraphics[width=\linewidth]{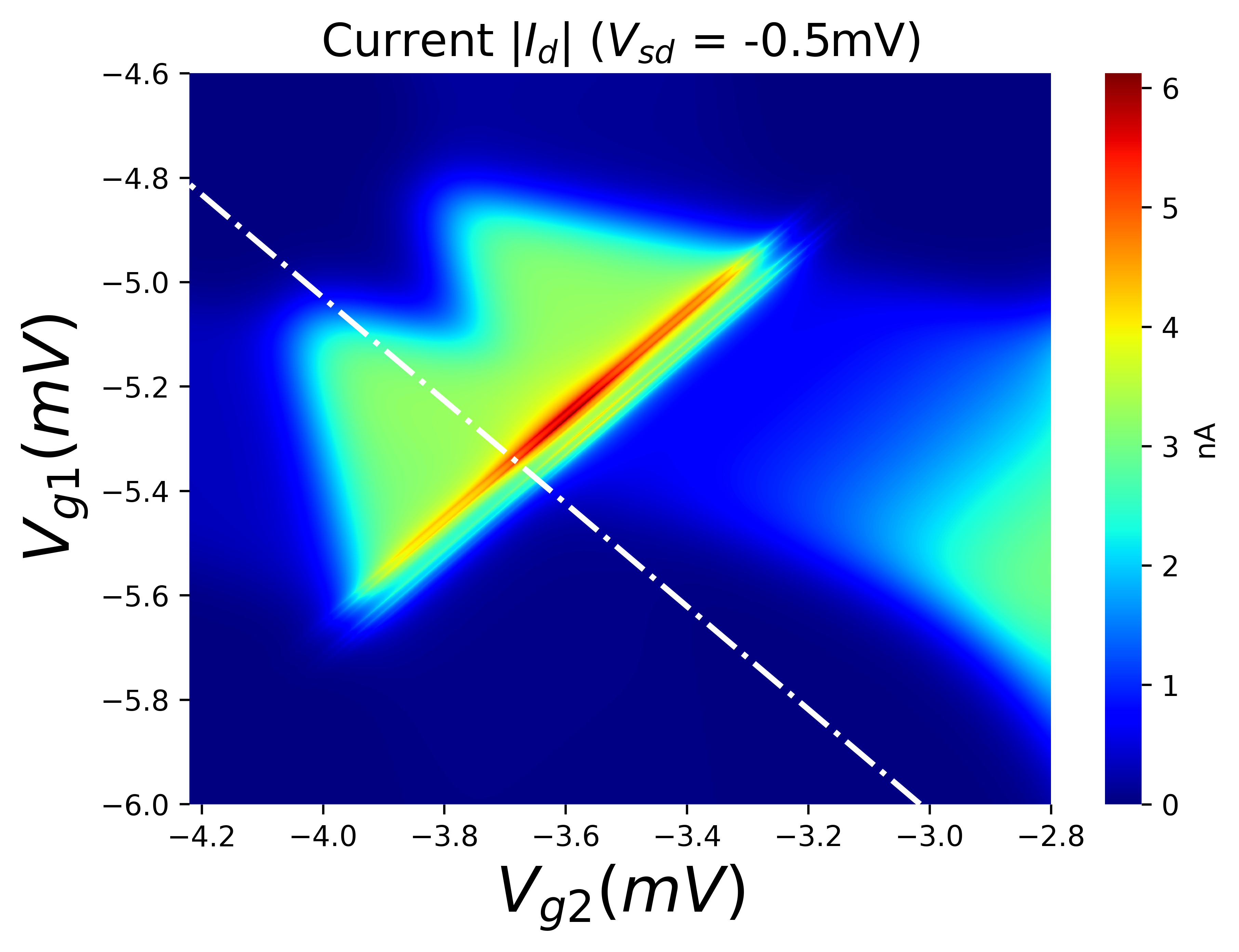}
    \includegraphics[width=\linewidth]{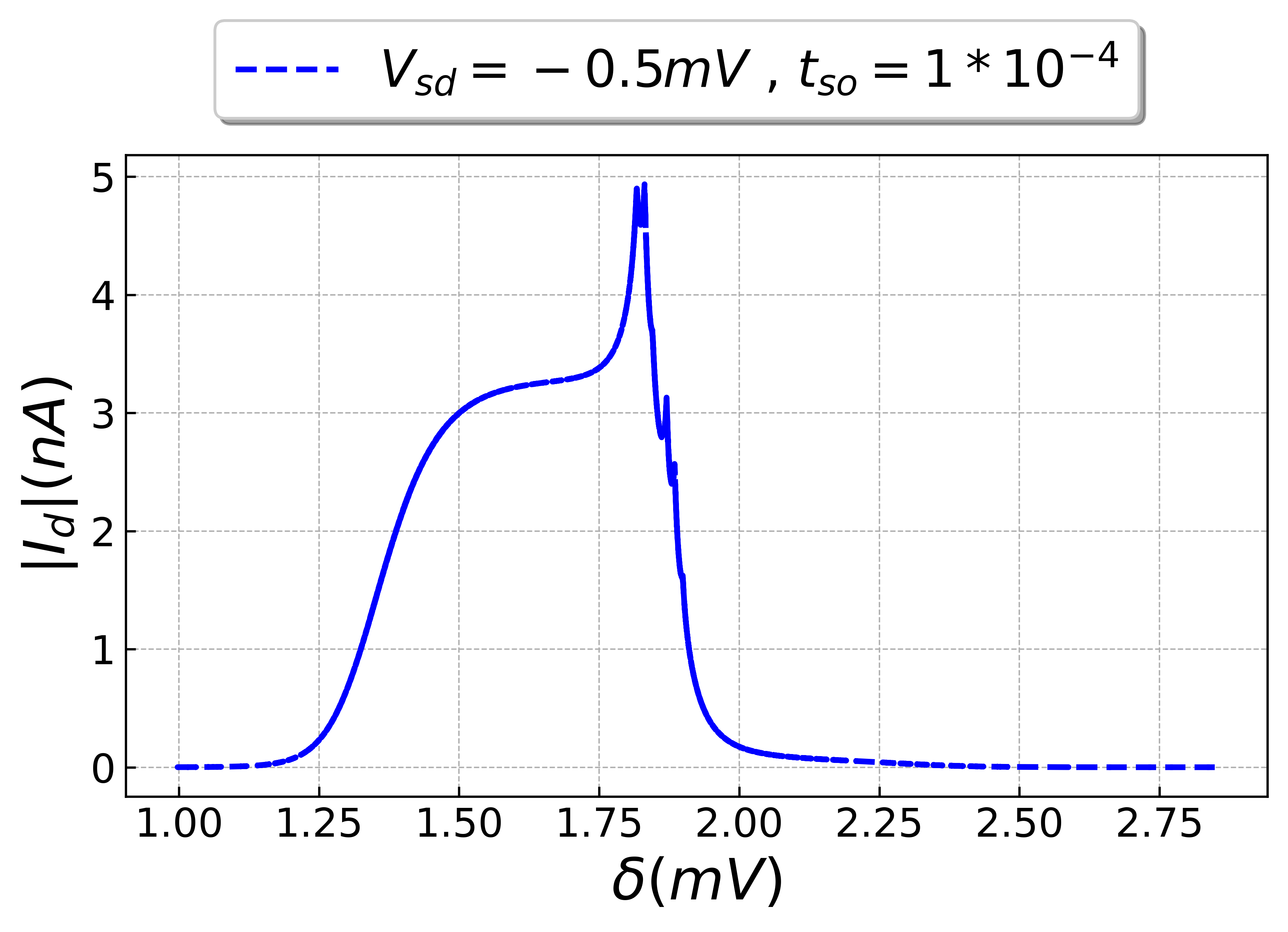}
    \label{fig:tso_ons}
\end{minipage}}
\subfloat[]{
\begin{minipage}{0.32\linewidth}
\centering
    \includegraphics[width=\linewidth]{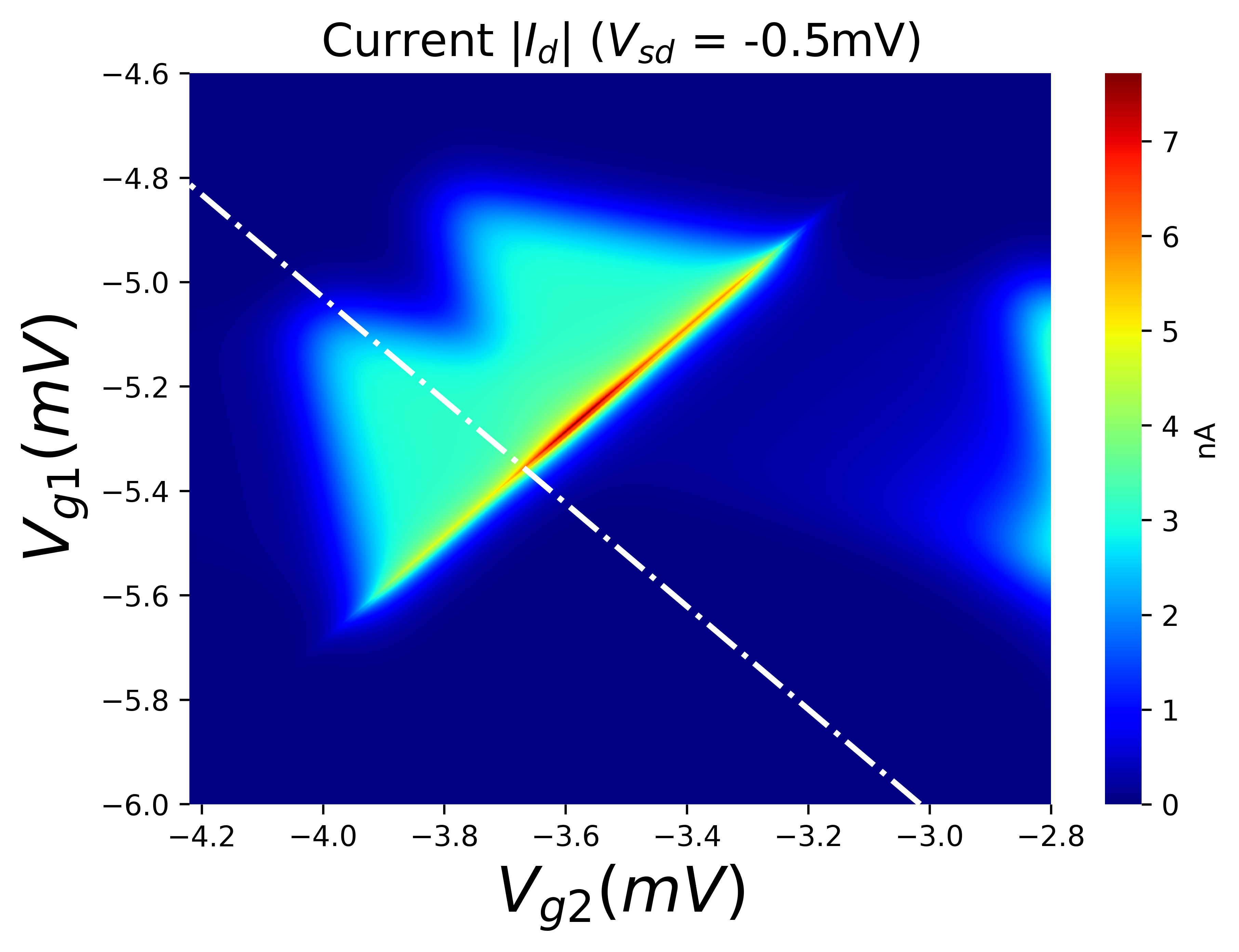}
    \includegraphics[width=\linewidth]{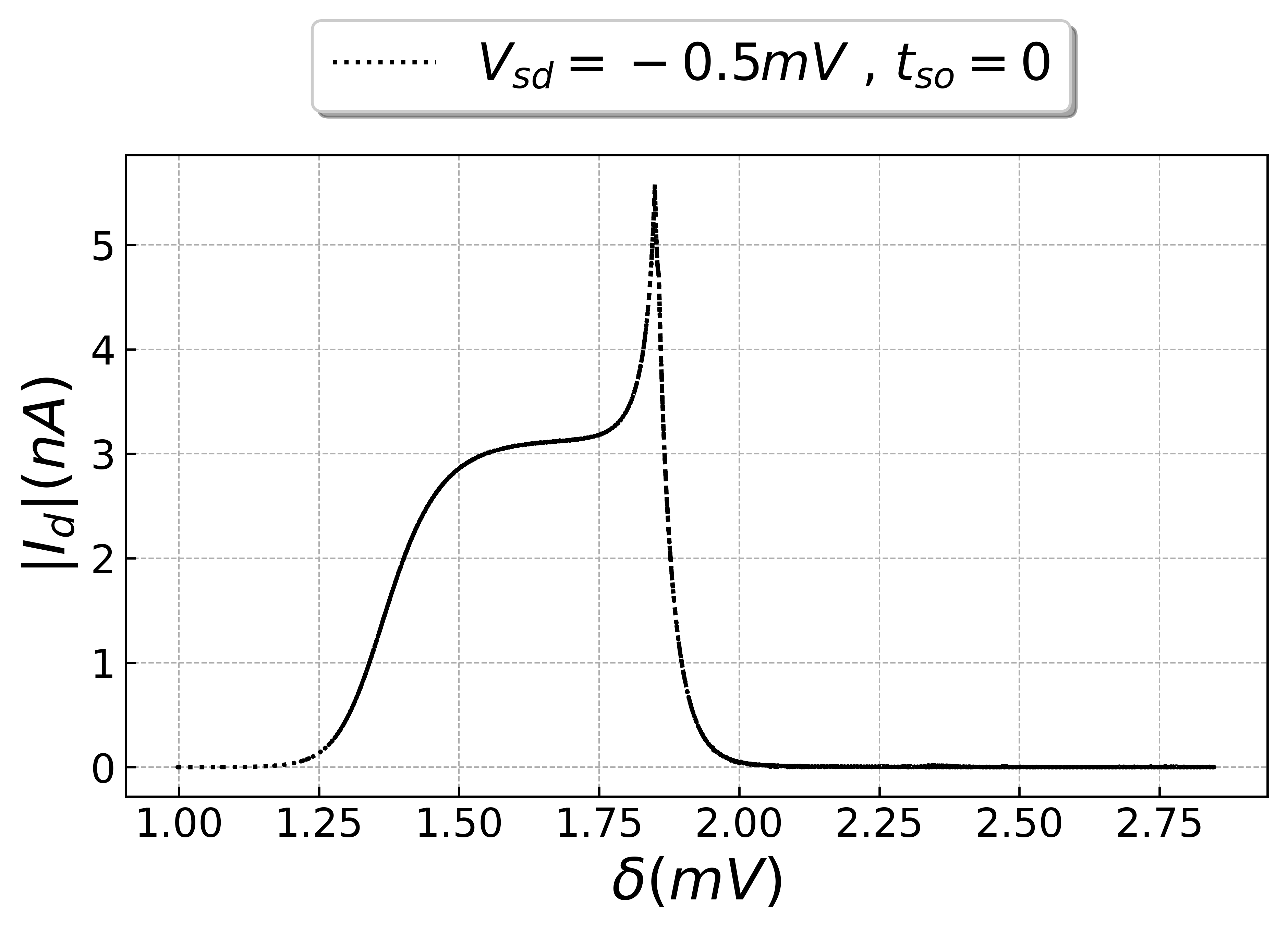}
    \label{fig:tso_offs}
\end{minipage}}
\subfloat[]{
\begin{minipage}{0.32\linewidth}
\centering
    \includegraphics[width=\linewidth]{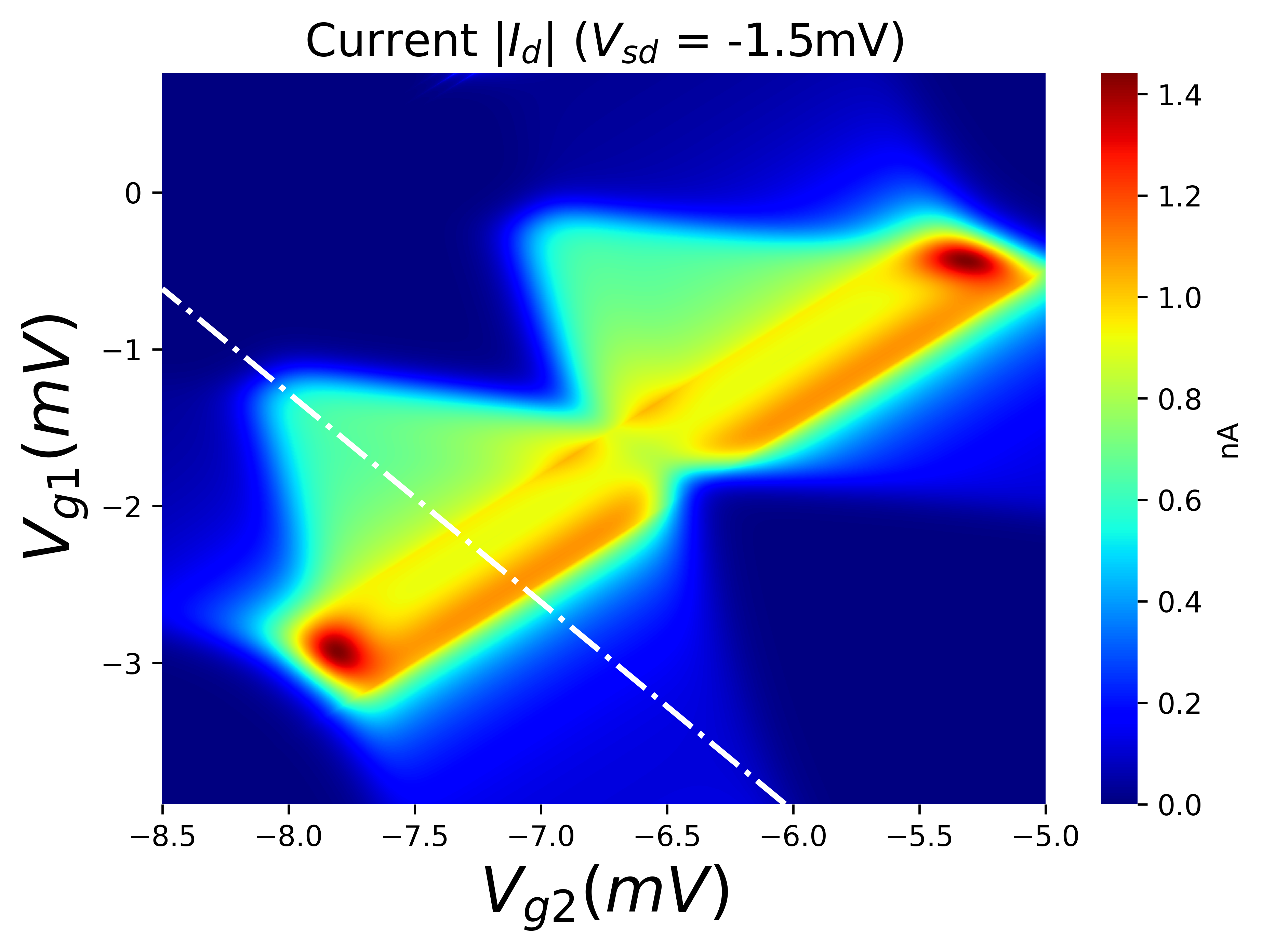}
    \includegraphics[width=\linewidth]{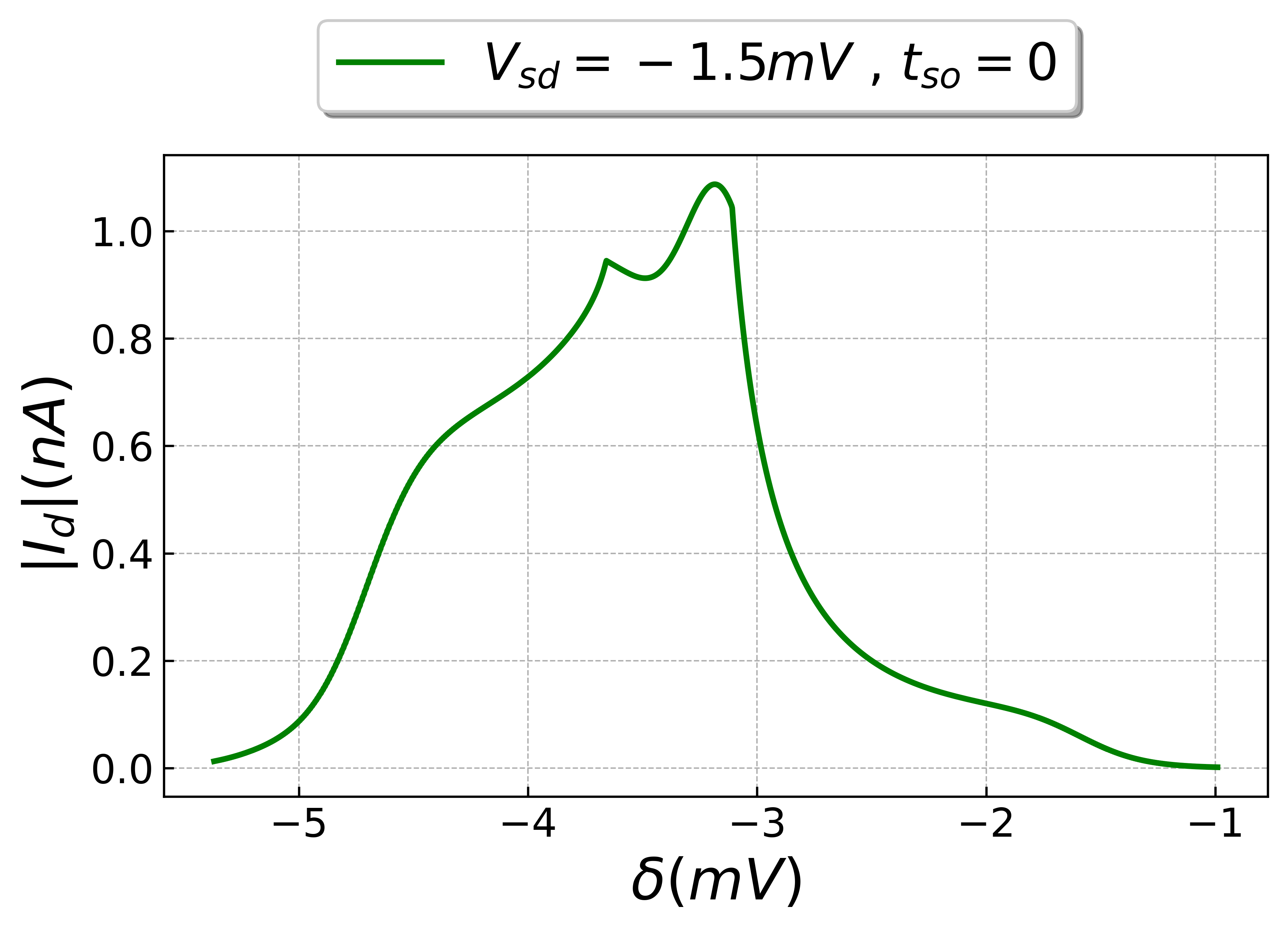}
    \label{fig:multi_blocks}
\end{minipage}}
%

%
\captionsetup{justification=raggedright,singlelinecheck=false}
\caption{
Occurrence of multiple excitations and multiple blockades in a single triangle.
\textbf{(a)} On including spin-flip tunnelling arising out of spin-orbit coupling ($t_{so}$) and applying a small parallel magnetic field ($B_{||} = 80 \mu T $), we observe new transitions, that were forbidden previously. These excited transitions lead to multiple resonances that manifest as multiple current peaks in the triangles.
\textbf{(b)} In the absence of spin-flip tunneling, the transitions happen between closely packed, almost degenerate states, showing no signatures of multiple excitations.
\textbf{(c)} A case of multiple blockades with two excitation peaks. 
Figures (d), (e), and (f) are cross-sections of the CS diagrams (a), (b), and (c), respectively. 
}
\label{fig:multi_BT}
\end{figure*}

\subsection{Transport formulation\label{subsec:formalism:transport}}
Despite having a solid grasp of the theory concerning current blockades in single-degree-of-freedom DQDs ~\cite{bhasky_main,bhasky_2,ono,benzene,dark_channels}, the behavior of the current in DQDs with multiple degrees of freedom has remained a challenge, until recently~\cite{Mukherjee2023}. The overall current in such a system arises from the intricate interplay between the probabilities of occupation for each eigenstate and the transition rates between them.
To address intricate problem, we expand upon the existing master equation, which is well-known in literature~\cite{bhasky_2}. The eigenstates are labelled $\ket{N,i}$, where $N$ denotes the total electron occupancy and $i$ denotes the $i^\text{th}$ state in the corresponding Fock state subspace with total electron occupancy $N$. We define the quantity $P^N_i$ to denote the probability of occupancy of the state $\ket{N,i}$ and $R^{L(R)}_{(N_1,i)\rightarrow(N_2,j)}$ to denote the rate of transition from the state $\ket{N_1,i}$ to the state $\ket{N_2,j}$ by injection or removal of an electron from the source(drain). Henceforth, we shall use the index $\alpha$ for the source ($\alpha=L$) or the drain ($\alpha=R$). The probabilities $P^N_i$ evolve over time as
\begin{align}
    \dot{P}^N_i &= \sum_j \left[R_{(N\pm1,j)\rightarrow(N,i)}P^{N\pm1}_j-R_{(N,i)\rightarrow(N\pm1,j)}P^N_i\right],\label{eq:master_equation}
\end{align}
where 
\begin{align}
    R_{(N_1,i)\rightarrow(N_2,j)}=\sum_{\alpha\in\{L,R\}}R^{\alpha}_{(N_1,i)\rightarrow(N_2,j)}.
\end{align}

\subsection{Inelastic processes\label{subsec:PAT}}
Besides resonant tunneling, inelastic processes and co-tunneling often play a pivotal role in determining the current-voltage characteristics. In fact, the origin of the base current in the bias triangles is attributed to inelastic processes\cite{spin_qubits, FUJISAWA2000413, BiasTri_exp}. To describe relaxation processes, an additional rate term $R^{(ph)}$ is introduced to cater to the bosonic degrees of freedom.
In a similar spirit of weak tunneling consideration, we consider weak coupling to the bosonic bath. Hence, the contributions from first order $\alpha_{ph}$ only contributes. The transition rates are given by
\begin{align}
        R_{(N_1,i)\rightarrow(N_1,j)}=\sum_{\alpha\in\{L,R\}}R^{\alpha(ph)}_{(N_1,i)\rightarrow(N_1,j)}.
\end{align}
We notice that the transitions take place among many-body states of the same manifold, i.e., within states with same occupancy number. Individual bosonic rates $R^{\alpha(ph)}_{(N_1,i)\rightarrow(N_1,j)}$ can be elaborated as
\begin{align}
        R^{\alpha(ph)}_{(N_1,i)\rightarrow(N_1,j)}=b(E)|\langle (N_1,i)| c^{\dagger} c |(N_1,j) \rangle|^2,
\end{align}
where $b(x) = sgn(x)\alpha_{ph}(x)n_b(x)$, with the Bose function $n_b(x) = \frac{1}{(exp(x/k_BT)-1)}$ and $c^{\dagger}$ represents corresponding creation operator.

\subsection{Machine learning methodology\label{subsec:formalism:machinelearning}}

Pauli blockades are essential for qubit initialisation and readout. We demonstrate an application of the above transport formalism to generate simulated data and train a deep learning network for automated detection of Pauli blockade. We follow an approach ~\cite{schuff2022identifying} using a residual convolutional neural network which is 18 layers deep~\cite{ResNet}. \\
\indent The training data set is generated by randomly varying multiple device parameters in the transport simulations. The simulations generate charge-stability plots (in terms of $I$ vs $V_{G1}$, $V_{G2}$) and a rectangular gate voltage window enclosing a pair of bias triangles is then extracted. These images are labelled as PB (Pauli blockade) and No PB (no Pauli blockade) according to the nature of their respective charge transitions. Bias triangles corresponding to positive and negative source-drain voltages are concatenated together and presented as training data to the deep learning classifier. The presence of Pauli blockade is indicated by the suppression of current in one bias direction as compared to the other. \\
\indent The machine learning model is implemented in PyTorch with CUDA. All the extracted images are converted to a grayscale format to simplify the training process and remove unnecessary colour information. A test-train split of 30\% is used and the simulated data is augmented by random stretching, rotation, shearing, and contrast changes. Additional augmentations from the Torchvision library including elastic transformation and sharpness adjustment are also applied to some inputs. Training is run for 20 epochs or till sufficient accuracy is reached, and a mini-batch size of 256 is used. The input data is downsized to about 68x60 pixels to make it similar to the resolution of typical current-voltage measurements. Softmax function is used to obtain respective probabilities for the two classes, Pauli blockade (PB) and no Pauli blockade (no PB). The sample is assigned to the class with higher predicted probability of the two (greater than 50\%). \\


\begin{figure*}[thpb]
\centering
%
\begin{minipage}{\linewidth}
\centering
    \includegraphics[width=0.8\textwidth]{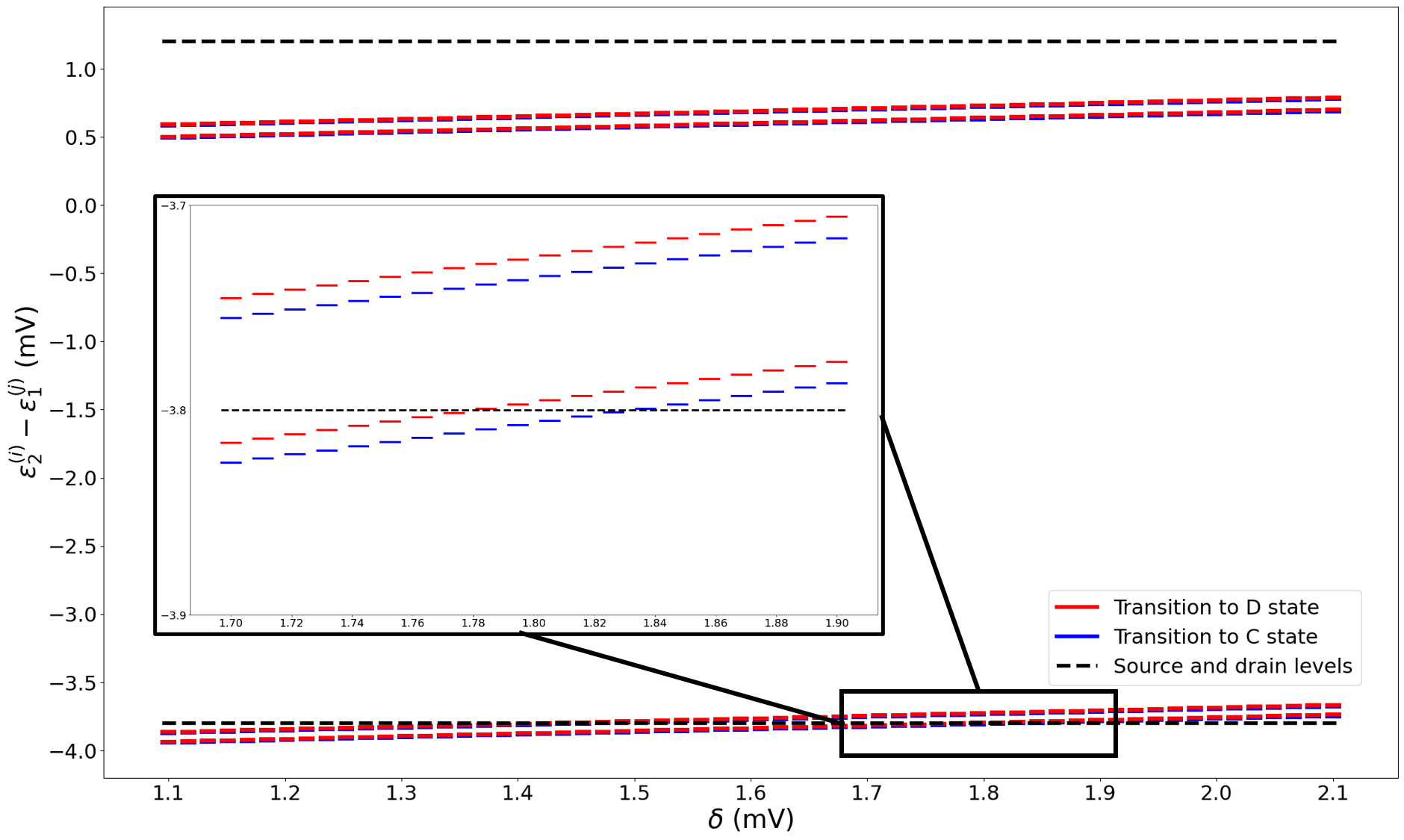}
\end{minipage}
\captionsetup{justification=raggedright,singlelinecheck=false}
\caption{
A plot of the energy difference between the $N=1$ and $N=2$ eigenstates for transitions allowed by selection rules. The plot is corresponding to the multiple excitations as shown in FIG~\ref{fig:tso_ons}. At a detuning of around $1.78$ meV, we observe the start of the first blockade, this is precisely where a $\mathcal{D}$ transition enters the source-drain bias window in the presence of a corresponding $\mathcal{C}$ transition. The inset shows a zoom in for detuning in the range of multiple peaks. There is another peak observed when a new $\mathcal{C}$ transition enters the bias window, at around $1.83$ meV.
}
\label{fig:e_det}
\end{figure*}

\section{Results}
\label{sec:results}
The charge stability diagram is an important illustrative in understanding the effect of transport dynamics in the system. It is essentially a current (conductance) map plotted with respect to the gate voltages used to tune the DQD potential barriers. We carry out a variety of analyses on the data that we obtain from our physics based simulations. Some of the primary parameters that play a crucial role in obtaining the conductance plots are (i) source-drain bias ($V_{sd}$), (ii) photon coupling strength $\alpha_{ph}$, and (iii) the tunneling coefficients $t_c$, $t_{so}$, and $t_{vo}$. The other tunable parameters kept fixed are the temperature $T_c$ (= 0.9K) and cross-capacitance C (= 0.2). \\
\indent We first discuss the dependence of bias triangles on primary system parameters as illustrated in Fig.~\ref{fig:Origin_BT}. Next, we describe the occurrence of multiple peaks and blockades in Fig.~\ref{fig:multi_BT}. Finally, we present some results from the use of deep learning on our simulated data for detecting the occurrence of Pauli blockade.
\subsection{Origin of the bias triangles}
Figures ~\ref{fig:high_b_cias} and \ref{fig:low_b_cias} show the difference in charge stability diagrams for the high and low bias regimes. In Fig.~\ref{fig:high_b_cias}, distinct triangles are visible when the bias is set high ($V_{sd} = -1.0 mV$). However, as depicted in Fig.~\ref{fig:low_b_cias}, we observe that the distinct bias triangles are not visible for low bias ($V_{sd} = -0.2 mV$). Initially, only a single transition (between ground states) is possible in the low-bias regime. Multiple conduction channels open up for electrons on increasing the bias, leading to conduction between $N-1$, $N$, and $N+1$ states, and the corresponding excitations as permitted via the transport selection rules \cite{Mukherjee2023}. \\
\label{subsec:results: blockade}
\indent We focus on transitions featuring the (0,2) and the (1, 1) configuration. At a certain bias of $V_{sd} = -1.0 mV$, Fig.~\ref{fig:for_pat} depicts the bias triangle with the current. The triangle's base corresponds to the ground state-ground state (GS-GS) transition between two different charge (number) configurations, leading to larger current values. On reversing the bias, as seen in Fig.~\ref{fig:rev_pat}, it can be noticed that there is about a three to four fold drop in the current values. The PAT processes play the role of inelastic scattering crucial in forming the triangle. If the boson coupling is turned off (Fig.~\ref{fig:rev_nopat}), we notice a further decrement in the current(about 100 times smaller) than in the non-blockaded case. A comparison of the blockaded and non-blockaded cases along a cut-line is presented in Fig.~\ref{fig:detunes}. In agreement with the theory, the GS-GS transitions stay the same in both forward and reverse biases with only a drop in magnitude.
\subsection{Multiple peaks and blockades}
The addition of spin-flip terms in the Hamiltonian leads to further splitting of the states giving rise to more resonant lines in the triangles, with a small spin-coupled magnetic field ($B_{||} = 80 \mu T $), and a finite tunneling coefficient ($t_{c}= 10^{-4}$). Figure~\ref{fig:tso_ons} depicts a case with the spin-flip present ($t_{so} = 10^{-4}$), while Fig.~\ref{fig:tso_offs} is the same triangle with the spin-flip term absent ($t_{so} = 0$). The contrast can be observed in that for the former case, there are many minute peaks near the base of the triangle, while for the latter, a single GS-GS transition is observed. \\
\indent Under a certain set conditions of $B_{\perp}$ (=3.2 T), we report the signature of multiple blockades. Figure~\ref{fig:multi_blocks} depicts one such case in which two peaks sandwich a large dip across the white cut line. It can be inferred that there are thus three multi-blockaded regions in the triangle. It is essential to note that though the multiple excitation cases and the multi-blockade cases discussed here seem similar in terms of the effect, the cause for their origin varies.\\
\indent Multiple Pauli blockades arise when specific conditions are met that lead to a negative differential resistance (NDR)~\cite{BM_2007,Mukherjee2023}. Multiple blockades arise because of the spin and valley Zeeman effects, wherein, the eigenergies in the Fock space depend upon the eigenstate, i.e., the degeneracy in energy between the $C$ and $D$ states is broken. The master equation then yields the relation between the energies, and in turn the gate voltages, that will lead to multiple Pauli blockades. \\
\indent Multiple excitations occur in the presence of effects such as strong spin-orbit coupling or photon-qubit interactions, which leads to multiple current peaks (resonances). Multiple resonances arise due to a marked difference in the transition energies between the eigenstates in the $N$ and $N+1$ occupancy Fock-states. In the presence of sequential tunneling alone, the eigenstates formed in the $N=2$ Fock space are of types $C$ or $D$ only. The coefficients within the $C$ or the $D$ state govern the occurence of multiple blockades. In contrast, spin-flip tunnelling results in the eigenstates being a superposition of two $C$ states or $D$ states. In such cases, the coefficients with which two $C$ (or $D$) states mix dictate where a resonance line can be observed. This leads to multiple resonance lines within a single bias traingle, unlike multiple bias triangles in case of pure sequential tunneling case. Figure~\ref{fig:e_det} demonstrates the relationship between the energy eigenspectrum and the occurrence of blockades.
\begin{figure*}[thpb]
\centering
%
\begin{minipage}{\linewidth}
\centering
    \includegraphics[width=\textwidth]{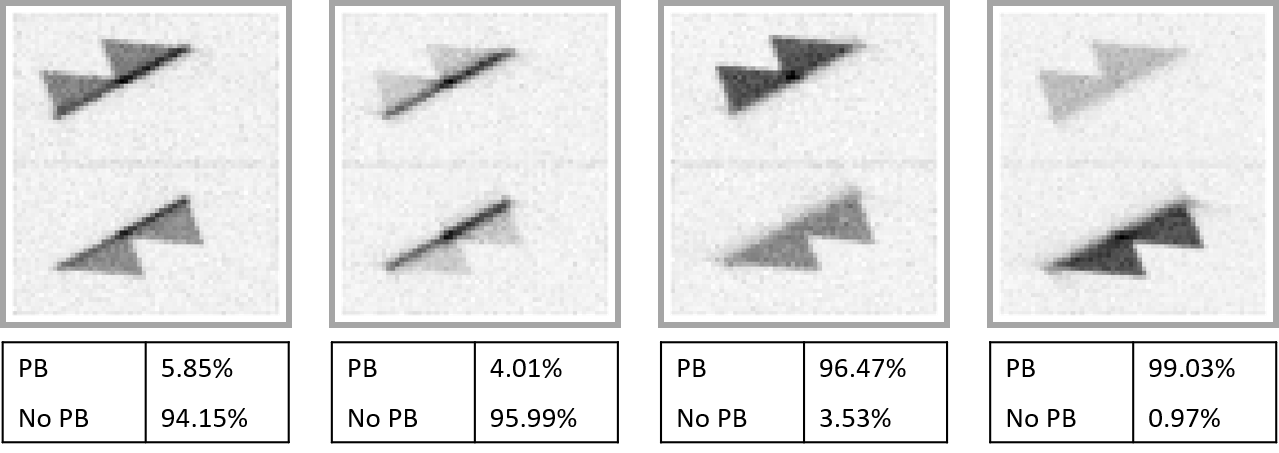}
\end{minipage}
\captionsetup{justification=raggedright,singlelinecheck=false}
\caption{
Predicted labels for the simulated test data using the deep neural network. The classifier takes in a concatenated pair of bias triangle images at positive and negative source-drain voltage for the same $V_{G1}$ and $V_{G2}$ range. The values below the pair of images indicate the probability predicted by the classifier for Pauli Blockade and no Pauli Blockade occurring in the given bias triangles.
}
\label{fig:ML1}
\end{figure*}
\subsection{Identification of Pauli blockades using machine learning}
Figure~\ref{fig:ML1} presents a few samples from the test data set along with labels and probabilities as predicted by the deep learning classifier. It can be seen that the lower image of the concatenated pair corresponds to a positive $V_{SD}$ and the upper image to a negative $V_{SD}$. The overall test accuracy obtained for our implementation is 96.15\%. Images in the figure reflect the actual input resolution to the classifier after augmenting and downsizing them. For the purpose of assigning predicted labels and calculating test accuracy, the sample is assigned to the class with higher probability, as mentioned before. \\
\indent As a way to benchmark the performance of our neural network trained on the the simulated data set, we also perform the classification on a limited set of data from experimental results which have appeared in the literature recently \cite{PauliBlockadeDataRep, main_exp_blg, tong2023pauli}. An accuracy of 94.44\% (17 out of 18 test cases predicted correctly) was obtained on the experimental test data set, thus showing the robustness of our neural network classifier and the simulated training data. This also re-validates the importance of using simulations in addition to experimental data for training, when the availability of experimental data is scarce \cite{schuff2022identifying, zwolak2018qflow, lennon2019efficiently, PhysRevApplied.13.034075}. \\
\begin{figure}[t]
    \centering
    \includegraphics[width=\linewidth]{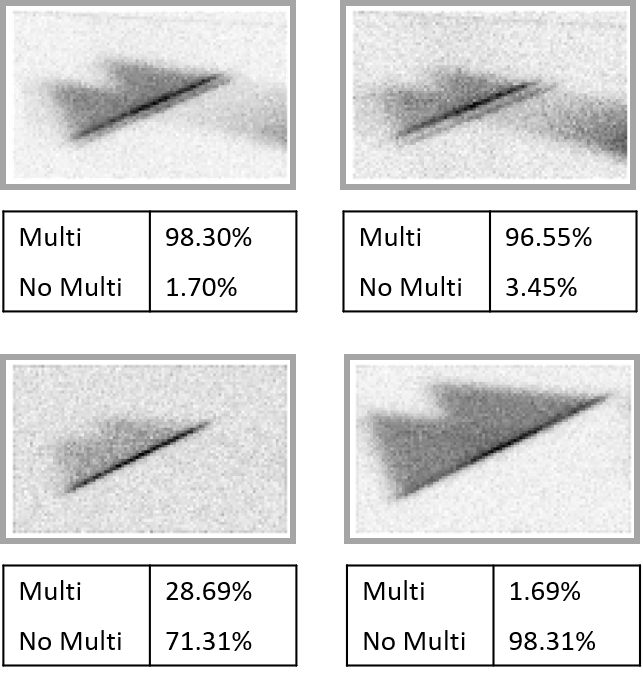}
    \caption{Detection of multiple peaks using the machine learning classifier. Predicted probability for the presence or absence of multiple peaks is given below the corresponding images.}
    \label{fig:ML2}
\end{figure}
\indent The machine learning procedure as mentioned above can also be used to detect and classify other kinds of features in the bias triangles, such as the occurrence of multiple peaks. This is depicted in Fig.~\ref{fig:ML2}, where the upper row represents bias triangles showing multiple peaks, while the images in the lower row only has a single GS-GS transition. The two classes are multiple peaks present (Multi) and multiple peaks not present (No Multi). The training procedure used in this case is similar to that for Pauli blockade detection and the test accuracy obtained on simulated data is 98.36\%.
\section{Conclusion}
\label{sec:conclusion}
Focusing on DQD structures in the BLG platform, and the experimental results available in literature, we first built theoretical models to capture the intricate interplay between externally fed gate voltages and the physical properties of the DQD setup, allowing us to effectively simulate Pauli blockades. Employing the master equations for transport and considering extrinsic factors such as electron-photon interactions, we thoroughly investigated all potential occurrences of Pauli blockades. Notably, our research revealed two remarkable phenomena: (i) the existence of multiple resonances within a bias triangle, and (ii) the occurrence of multiple Pauli blockades. Leveraging our model to train a machine learning algorithm, we successfully developed an automated method for real-time detection of multiple Pauli blockade regimes. Through numerical predictions and validations against test data, we identified where and how many Pauli blockades are likely to occur. We propose that our model can effectively detect the generic class of Pauli blockades in practical experimental setups and hence serves as the foundation for future experiments on qubits that utilize 2-D material platforms.
\section*{Acknowledgements}
 The author BM acknowledges the support by the Science and Engineering Research Board (SERB), Government of India, Grant No. CRG/2021/003102 and Grant No. MTR/2021/000388.

\bibliography{main2}

\end{document}